\documentclass[12pt]{article}

\usepackage{latexsym,amsmath,amssymb,theorem,epsfig}

\DeclareFontFamily{OT1}{rsfs10}{}
\DeclareFontShape{OT1}{rsfs10}{m}{n}{ <-> rsfs10 }{}
\DeclareMathAlphabet{\mathscript}{OT1}{rsfs10}{m}{n}

\numberwithin{equation}{section}

\newcommand{\ns}{\normalsize}

\def\gsim{ \lower .75ex \hbox{$\sim$} \llap{\raise .27ex \hbox{$>$}} }
\def\lsim{ \lower .75ex \hbox{$\sim$} \llap{\raise .27ex \hbox{$<$}} }
\def\be{\begin{equation}}
\def\ee{\end{equation}}
\def\bea{\begin{eqnarray}}
\def\eea{\end{eqnarray}}

\theoremstyle{plain}

{\theorembodyfont{\rmfamily} }

\begin{document}


\begin{titlepage}

\vspace{-5cm}

\title{
  \hfill{\ns }  \\[1em]
   {\LARGE Dynamically SUSY Breaking SQCD on F-Theory Seven-Branes}
\\[1em] }
\author{
   Evgeny I. Buchbinder 
     \\[0.5em]
   {\ns Perimeter Institute for Theoretical Physics} \\[-0.4cm]
{\ns Waterloo, Ontario, N2L 2Y5, Canada}}

\date{}

\maketitle

\begin{abstract}

We study how dynamically breaking SQCD can be obtained on 
two intersecting seven-branes in F-theory.
In the mechanism which we present in this paper 
one of the seven-branes is responsible for producing the low-energy
gauge group and the other one is for generating vector bundle moduli. 
The fundamental matter charged under the gauge group is localized 
on the intersection. The mass of the matter fields is controlled by the vector bundle 
moduli. The analysis of under what conditions a sufficient number 
of the fundamental flavors
becomes light turns out to be equivalent to the analysis 
of non-perturbative superpotentials for vector bundle moduli in Heterotic M-theory. 
We give an example in which 
we present an explicit equation in the moduli space whose 
zero locus corresponds to the fundamental fields becoming light. 
This allows us to provide a local F-theory realization of 
massive ${\cal N}=1$, $SU(N_c)$ SQCD in the 
free magnetic range which dynamically breaks supersymmetry. 

\end{abstract}

\thispagestyle{empty}

\end{titlepage}


\section{Introduction}


F-theory~\cite{Vafa} compactified to four dimensions is potentially one of 
the most promising ways to obtain phenomenological models.
One of the attractive features of F-theory is that,
unlike most of the intersecting brane models, it naturally provides 
Grand Unification at the compactification scale. In addition, 
F-theory compactifications admit a rich structure of branes and fluxes 
which suggests a potential variety of possibilities to build   
quasi-realistic phenomenology. 

One of the reasons to expect interesting particle physics in F-theory is 
due to its relation with heterotic string. For a certain type 
of heterotic compactifications, namely on elliptically fibered Calabi-Yau 
manifolds, the heterotic/F-theory duality is relatively well 
understood~\cite{MVafa1, Mvafa2, Vafasix, FMW, Tony, Curio, Andreas1, Andreas2, Diaconescu0}. 
Since heterotic compactifications are known to naturally lead to quasi-realistic 
phenomenological models one should expect the same on the F-theory side. 
Recently, vacua with the spectrum of the supersymmetric standard model 
were obtained in heterotic compactifications on non-simply connected 
Calabi-Yau manifolds~\cite{Braun1, Braun2, Braun3, Braun4, Ron1, Ron2}.
The relation between heterotic string and F-theory suggests 
that such models can probably also be found on the F-theory side. 
On the other hand, the general structure of F-theory compactifications has certain 
advantages comparing to that in heterotic string theory. 
The particle sector in F-theory is localized on (in general intersecting)
seven-branes. It implies that in order to study particle physics in F-theory
one is likely to need to know only the local structure of the F-theory Calabi-Yau 
manifold near the seven-branes. On the contrary, on the heterotic side, 
there are no branes involved in the model building and it is not possible 
to reduce the problem to a local consideration. Another attractive 
feature of F-theory, or type IIB compactifications in general, is a recent progress
in moduli srabilization (see~\cite{DougK, Denef1} for a review)
and cosmological applications (see~\cite{Baumann, Lust0} for the most recent reports).

On the other hand, the particle spectrum of the F-theory compactifications 
is very poorly understood and its study represents a difficult problem. 
It is hard to approach this problem from the type IIB string theory side because
F-theory compactifications are intrinsically non-perturbative. 
Away from certain orientifold limits one cannot obtain the spectrum in a simple way
by quantizing open strings ending on the F-theory seven-branes.
The approach based on using duality with heterotic string theory 
is also problematic since it is not known how the duality map acts on the heterotic 
spectrum. In general, it is a complicated mathematical problem.
A progress in this direction has recently been reported in~\cite{Martijn, BHV}
(see also~\cite{Tatar}).
In particular, in~\cite{BHV}, Beasley, Heckman and Vafa constructed a field
theory on intersecting seven-branes. The approach of~\cite{BHV} was to start 
with the maximally supersymmetric Yang-Mills theory and twist it in such a way that 
the theory on the branes preserves only four supercharges. 
The authors of~\cite{BHV} showed that such a twist is unique. Once the theory 
on the seven-branes is known one can study the particle spectrum in four dimensions
just like in heterotic compactifications. The analysis in~\cite{BHV} relies
only on the local geometry near the seven-branes. However, one can expect that 
it rather adequately describes the particle sector of F-theory compactifications
though global restrictions in some cases can be important. 

The goal of~\cite{BHV, BHV2} is 
to study GUT theories in the F-theory framework. 
However, in general,
it is interesting to look not only at quasi-realistic theories in the visible sector
but also at hidden sectors. 
One of the important questions in string theory model building 
is how supersymmetry can be broken in these models. A natural 
attempt would be to create a hidden sector which breaks supersymmetry
and then to communicate this breaking to the visible sector 
via some mediation mechanism. The most recent progress on dynamical 
SUSY breaking was achieved by Intrilligator, Seiberg and Shih in~\cite{ISS} 
where it was shown that a class of ${\cal N}=1$ SQCD theories has a metastable 
SUSY breaking vacuum at strong coupling. This class involves theories
whose matter spectrum consists of $N_f$ massive fundamental flavors
where $N_f$ is in the free magnetic range. 
It is important to understand how field theories dynamically breaking SUSY 
in the infrared, like the one studied in~\cite{ISS}, can be embedded
in realistic string compactifications with stable moduli. 
This has been discussed in various contexts in~\cite{Diaconescu, BBO1, BBO2}. 

In this paper, we will discuss how the field theory model of~\cite{ISS}, 
that is massive SQCD in the free magnetic range can be obtained on the seven-branes
of F-theory. More precisely, we consider the theory on two 
intersecting seven-branes. The field theory action
for this system was obtained in~\cite{BHV}. Since we have a four-dimensional 
theory, the seven-branes wrap two different complex surfaces and extend in four dimensions.
In addition, they intersect along a complex curve. 
All these objects, namely
the two seven-branes and the curve play an important role in our construction. 
The theory on one of the seven-branes is pure ${\cal N}=1$, $SU(N_c)$ Yang-Mills theory 
without matter. The role of the second seven-brane is to contribute vector 
bundle moduli. To achieve it, we put a non-trivial instanton on the surface 
which this seven-brane wraps. The matter in the (anti)-fundamental representation 
of the gauge group $SU(N_c)$ comes from the intersection curve. In order to generate 
the field theory of~\cite{ISS} the matter has to receive a relatively small mass.
In global compactifications, there are no free constant parameters which can be used 
for this purpose. 
The role of parameters is played by moduli fields which have to be stabilized in any 
quasi-realistic compactification. In our case, the relevant moduli are the moduli of the vector 
bundle. The mass of the fundamental fields localized on the intersection curve 
is a function of these moduli. In fact, in a generic point in the moduli space, 
all the matter fields are very massive and have to be integrated out at low energies.
The resulting theory in this case is $SU(N_c)$ supersymmteric Yang-Mills theory.
However, near some special subvarieties in the moduli space a certain number 
of the fundamental fields can become light and the resulting theory 
is SQCD with slightly massive flavors. If one can control how many fundamentals become
light near various subvarieties in the moduli space, one can obtain massive SQCD 
in the free magnetic range. We show that this problem of analyzing under what conditions
there is light fundamental matter is exactly equivalent to the problem of computing  
non-perturbative superpotentials for vector bundle moduli in 
Heterotic M-theory~\cite
{Greene, Witten99, BDO2, BDO3}. The holomorphic
function which was the superpotential in the Heterotic M-theory context 
now defines the subvariety near which there are light fundamental fields. 
To have an analytic control over the problem, 
we choose one of the surfaces to be the rational elliptic surface $d P_9$. 
This surface admits an elliptic fibration so that we can use the
spectral cover construction~\cite{FMW, Donagi} to build an instanton on it. 
As the result, we can write an explicit equation in the moduli space 
which governs the appearance of light fundamental fields as well as their number.
More precisely, the fundamental flavors parametrize the kernel of a certain square matrix. 
Therefore, the number light flavors coincides with the amount 
by which the rank of this matrix drops as we move in the locus 
of the zero determinant. 

This paper is organized as follows. In section 2, we give a review of F-theory 
compactifications. In particular, we review the field theory on intersecting
seven-branes constructed in~\cite{BHV} with focus on how the 
four-dimensional particle spectrum is encoded in the geometry of the branes. 
In subsection 3.1, we state, following~\cite{ISS},
the criteria that field theories
with dynamical supersymmetry breaking must satisfy. 
In the rest of section 3, we study how such theories can be embedded 
in F-theory. In section 4, we present a concrete realization of the ideas developed 
in section 3. 
We give an example where the holomorphic function in the moduli space,
near the zero locus 
of which we obtain massless fundamental matter, can be explicitly derived. 
We analyze how many fundamental flavors become light 
in different regimes in the moduli space
and give examples of SQCD in the free magnetic range. 
In addition, we show that the mechanism of generating light fields 
as we move in the vector bundle moduli space is precisely equivalent 
to having a Yukawa-type superpotential in the Lagrangian.
This superpotential is quadratic in the matter fields with the mass matrix depending 
on the vector bundle moduli. 
In conclusion, we briefly summarize our results and
discuss a possible extension of this work. 
Finally, Appendices A, B and C are devoted to some technical details. 


\section{F-Theory Compactifications}


\subsection{The General Structure}


In this section, we will review the structure of F-theory 
compactifications~\cite{Vafa, MVafa1, Mvafa2, Vafasix, FMW, Tony}. 
We will start with a general consideration and then review details 
of the field theory on the seven-branes~\cite{BHV}. 

F-theory is a special class of supersymmetric type IIB string compactifications
on a manifold which we will denote $Y$. This compactification has a non-trivial 
holomorphic axion-dilation which varies along $Y$. It can become singular 
and undergo an $SL(2, {\mathbb Z})$ monodromy along some divisor in $Y$ which we 
will denote $\Delta$. This can be interpreted as a compactification on a Calabi-Yau 
manifold $X$ which is elliptically fibered over $Y$ with $\Delta$ being the discriminant 
divisor of the the elliptic fibration over which the fibers degenerate. 
The position of $\Delta$ is interpreted as the location of the seven-branes on which 
the particle sector of the compactification is localized. The gauge group is determined 
by the type of the singularity along $\Delta$. 
In this paper, we will be interested in compactifications to four dimensions. Then 
$X$ is a Calabi-Yau four-fold, elliptically fibered over $Y$, 
and $\Delta$ is a surface in $Y$. In many cases, $\Delta$ is reducible and
has irreducible components intersection along a curve or points. In such situations, 
the particle sector can be viewed as an intersecting brane model where the seven-branes
wrap surfaces in the four-fold $X$ and extend in the four non-compact dimensions. 

To provide a global realization of such Calabi-Yau four-folds is a rather complicated task.
However, there is a class of F-theory compactifications whose global properties are relatively
well understood. These are F-theory $n$-fold compactifications dual to 
heterotic compactifications on an elliptically fibered Calabi-Yau $(n-1)$-fold
with a vector bundle whose structure group is in 
$E_8 \times E_8$.~\footnote{There is additional data involved in this duality.
The vector bundle on the heterotic side has to be constructed using an {\it irreducible} 
spectral cover. For simplicity, we will omit these details.}
Let us give a brief review of the Calabi-Yau manifold
$X$ is this case. 
This will provide us with some intuition about the general structure of the 
F-theory compactifications which will be used to motivate 
some of the choices we make further in the paper.
For concreteness, we will discuss the case $n=4$ which corresponds to compactifications
to four dimensions on both sides of the duality. In this case, 
the F-theory four-fold is described by a Weierstrass model 
\begin{equation}
y^2=x^3+ f(z_0; z_1, z_2)x+ g(z_0; z_1, z_2). 
\label{1.1.1}
\end{equation}
For fixed $(z_0; z_1, z_2)$ this equation describes an elliptic fiber.
The coordinate $z_0$ parametrizes ${\mathbb P}^1$ and $f(z_0; z_1, z_2)$ and 
$g(z_0; z_1, z_2)$ are polynomials of degree eight and twelve in $z_0$ respectively
\begin{eqnarray}
&&f(z_0; z_1, z_2)=\sum_{a=0}^8 f_a (z_1, z_2) z_0^a, \nonumber \\
&&g(z_0; z_1, z_2)=\sum_{b=0}^{12} f_b (z_1, z_2) z_0^b. 
\label{1.1.2}
\end{eqnarray}
For fixed $(z_1, z_2)$ eqs.~\eqref{1.1.1}, \eqref{1.1.2} describe elliptically fibered $K3$ surface. 
Indeed, the discriminant of the fibration given by 
\begin{equation}
\Delta=4f^3 + 27 g^2
\label{1.1.3}
\end{equation}
is a polynomial of degree 24 in $z_0$ meaning that the fiber degenerates over 24 point is 
${\mathbb P}^1$. Thus, $X$ is also a $K3$ fibration over a complex two-dimensional manifold 
parametrized by $(z_1, z_2)$. This space is identified with the base $B$ of the elliptically 
fibered three-fold on the heterotic side. 
In other words, elliptically fibered Calabi-Yau threefold with base $B$ on the 
heterotic side is mapped by duality to the F-theory elliptically fibered 
Calabi-Yau four-fold $X$ given by a Weierstrass model~\eqref{1.1.1} 
which is also $K3$ fibered over $B$. It was shown that the middle coefficients
$f_4(z_1, z_2)$ and $g_6(z_1, z_2)$ in eqs.~\eqref{1.1.2} encode the information about the 
complex structure of the heterotic threefold. The coefficients 
$f_a, a=0, \dots, 3$ and $g_b$, $b=0, \dots, 6$ encode the information about one of the 
$E_8$ vector bundles. Similarly, the coefficients
$f_a$, $a=5, \dots, 8$ and $g_b$, $b=7, \dots, 12$ describe 
the data of the other $E_8$ vector bundle.
Thus, to describe one of the particle sectors (visible or hidden) 
one can set the data of the second vector bundle to zero. Then $f(z_0; z_1, z_2)$ 
can be taken to be a polynomial of degree four in $z_0$ and $g(z_0; z_1, z_2)$
can be taken to be a polynomial of degree six. 

Let us now review the structure of the seven-branes. It is determined by the equation 
$\Delta=0$. The low-energy gauge group is determined by the singularity 
along the zero locus of $\Delta$ and is of the $ADE$-type.
All possible consistent singularities were 
obtained in~\cite{Vafasix} using the Tate's algorithm. In most cases, the discriminant 
divisor consists of two components intersecting along the curve $z_0=0$ which is just the base 
of the $K3$-fibration $B$. For example, the $E_7$ low-energy gauge group is 
described by the following coefficients $f(z_0; z_1, z_2)$ and $g(z_0; z_1, z_2)$ 
\begin{eqnarray}
&&f(z_0; z_1, z_2)=f_4 (z_1, z_2)z_0^4+ f_3 (z_1, z_2)z_0^3, \nonumber\\
&&g(z_0; z_1, z_2)=g_6 (z_1, z_2)z_0^6+ g_5 (z_1, z_2)z_0^5. 
\label{1.1.4}
\end{eqnarray}
The discriminant divisor can be obtained from eq.~\eqref{1.1.3} and is given by the 
zero locus of the following polynomial 
\begin{equation}
\Delta =z_0^9 (4 f_3^3 +12 f_3^2 f_4 z_0+ (27 g_5^2+ 54 g_5 g_6)z_0^2 +
(4 f_4^3 + 27 g_6^2) z_0^3). 
\label{1.1.4.1}
\end{equation}
We see that the discriminant divisor has two components. One is given by $z_0=0$ and the other one 
given by 
\begin{equation}
4 f_3^3 +12 f_3^2 f_4 z_0+ (27 g_5^2+ 54 g_5 g_6)z_0^2 +
(4 f_4^3 + 27 g_6^2) z_0^3 =0.
\label{1.1.5}
\end{equation}
These two surfaces intersect along the curve $f_3 (z_1, z_2)=0$. 
This and all other possible $ADE$-singularities were studied in detail in~\cite{Vafasix}. 


\subsection{The Theory on the Intersecting Seven-Branes}


To describe the particle spectrum of F-theory compactifications one has to study the
theory on the seven-branes. This analysis was performed in~\cite{BHV}. 
Motivated by the structure of the seven-branes in the globally known examples of F-theory
reviewed in the previous subsection,
we will concentrate on the models in which the discriminant
divisor consists of two smooth irreducible surfaces ${\cal S}$ and ${\cal S}^{\prime}$
intersecting along a curve ${\Sigma}$ which, for simplicity, we will assume 
to be smooth and irreducible. Let the singularities along ${\cal S}$ and ${\cal S}^{\prime}$ 
be of the type $G_{{\cal S}}$ and $G_{{\cal S}^{\prime}}$ respectively. Both 
$G_{{\cal S}}$ and $G_{{\cal S}^{\prime}}$ are of the $ADE$-type. This corresponds to 
$G_{{\cal S}}$ and $G_{{\cal S}^{\prime}}$ gauge groups on the two intersecting seven-branes.
We should note that 
one can also have a situation when there is no singularity along ${\cal S}$ or 
${\cal S}^{\prime}$. In this case, the gauge group on the world-volume of 
the corresponding seven-brane
is $U(1)$. 

To describe the theory on a seven-brane, the authors of~\cite{BHV} started with 
the maximally supersymmetric gauge theory on ${\mathbb R}^{1, 3} \times {\mathbb C}^2$. 
Then they replaced ${\mathbb C}^2$ with the component of the discriminant surface ${\cal S}$. 
The theory on ${\mathbb R}^{1, 3}\times {\cal S}$ has to preserve four supercharges. It was shown 
in~\cite{BHV} that this can achieved if one twists the maximally supersymmetric theory 
on ${\mathbb R}^{1, 3} \times {\mathbb C}^2$. Furthermore, it was shown in~\cite{BHV}
that there exists a unique twists preserving four supercharges. 
In a similar manner, one can analyze the theory on
${\mathbb R}^{1, 3} \times \Sigma$~\cite{BHV}. 
To make the paper self-contained, we give a review of the twisting in Appendices A and B. 

The resulting
action of the theory on the intersecting 
seven-branes is 
\begin{equation}
I=I_{{\cal S}}+I_{{\cal S}^{\prime}}+I_{\Sigma}.
\label{1.2.1}
\end{equation}
Here $I_{{\cal S}}$ is the action localized on ${\mathbb R}^{1, 3} \times {\cal S}$, 
$I_{{\cal S}^{\prime}}$ is the action localized on 
${\mathbb R}^{1, 3} \times {\cal S}^{\prime}$ and $I_{\Sigma}$ is the action localized 
on the intersection ${\mathbb R}^{1, 3} \times \Sigma$.
The precise form of $I_{{\cal S}}, I_{{\cal S}^{\prime}}$ and $I_{\Sigma}$ can be found 
in~\cite{BHV}. Here, we will only review the field content. 

Let us start with the fields localized on ${\mathbb R}^{1, 3} \times {\cal S}$. 
The first set of fields is 
\begin{equation}
(A_{\mu}, \eta_{\alpha}, \bar\eta_{\dot\alpha}), \quad \mu=0, \dots 3, \quad \alpha=1, 2
\label{1.2.2}
\end{equation}
which can be viewed as the vector multiplet. Here $A_{\mu}$ is the four-dimensional 
part of the $G_{{\cal S}}$-gauge field propagating on ${\mathbb R}^{1, 3}\times {\cal S}$.
Furthermore, $\eta_{\alpha}$ is a positive chirality spinor from the viewpoint of the 
four-dimensional Lorentz group. It also transforms in the adjoint representation 
of $G_{{\cal S}}$ (more precisely, it is a section of the adjoint bundle on  
${\mathbb R}^{1, 3}\times {\cal S}$). So far, all the fields in~\eqref{1.2.2} depend 
on the coordinates on ${\mathbb R}^{1, 3}$ as well on the coordinates on ${\cal S}$. 
The additional field can be viewed as (anti)-chiral multiplets
\begin{equation}
(A_m, \bar \psi_{\dot\alpha m}), \quad (A_{\bar m}, \psi_{\alpha \bar m})
\label{1.2.3}
\end{equation}
and 
\begin{equation}
(\phi_{m n}, \chi_{\alpha m n}), \quad (\bar \phi_{{\bar m} {\bar n}}, 
\bar \chi_{\dot\alpha {\bar m }{\bar n}}), 
\label{1.2.4}
\end{equation}
where $m, n=1, 2$, is the holomorphic index on ${\cal S}$. The fermions 
$\bar \psi_{\dot\alpha m}$ and $\chi_{\alpha m n}$, in addition to being 
sections of the adjoint bundle, transform as sections of 
$\Omega_{\bar\partial}^{(1, 0)}$ and $\Omega_{\bar\partial}^{(2, 0)}$
respectively. All field in eqs.~\eqref{1.2.3} and~\eqref{1.2.4} depend on the coordinates
on ${\mathbb R}^{1, 3}$ as well on the coordinates on ${\cal S}$.
To obtain the low-energy field theory in four-dimensions we compactify the action 
$I_{\cal S}$ on ${\cal S}$ and keep only the zero modes. To preserve supersymmetry, 
we have to satisfy some BPS conditions on ${\cal S}$. These conditions are as follows~\cite{BHV}
\begin{eqnarray}
&&F_{m n}=F_{{\bar m}{\bar n}}=0, \nonumber \\
&&\bar\partial_{A_m} \phi=0, \nonumber \\
&&\omega \wedge F+ \frac{i}{2}[\phi, \bar \phi]=0.
\label{1.2.5}
\end{eqnarray}
Here $F$ is the field strength constructed out of $(A_m, A_{\bar m})$. 
It can be viewed as a curvature of some vector bundle on ${\cal S}$. 
Furthermore, $\phi=\phi_{m n}d s^m d s^n$ is an adjoint-valued two-form on ${\cal S}$, 
$\bar\partial_{A_m}$ is the antiholomorphic covariant derivative and $\omega$ 
is the Kahler form. For simplicity, we will consider vacua with $\phi=0$. Then the equations
for $F$ become 
\begin{equation}
F_{m n}=0, \quad F_{{\bar m} {\bar n}}=0, \quad g^{m \bar n} F_{m \bar n}=0
\label{1.2.6}
\end{equation}
which are the Hermitian Yang-Mills equations on ${\cal S}$. This means that $F$ is the curvature 
on a stable holomorphic vector bundle on ${\cal S}$. Let $H_{{\cal S}}$ be the structure 
group of this vector bundle. Then in four dimensions $G_{{\cal S}}$ is broken 
to $\Gamma_{{\cal S}}$ which is the commutant of $H_{{\cal S}}$ in $G_{{\cal S}}$. 
Thus, after compactifying on ${\cal S}$ the action $I_{\cal S}$ is the action 
of the ${\cal N}=1$ supersymmetric gauge theory with gauge group $\Gamma_{{\cal S}}$ 
coupled to some matter. To obtain the matter content, we first 
decompose ${\rm ad} G_{{\cal S}}$ 
into irreducible representations of $\Gamma_{{\cal S}} \times H_{{\cal S}}$
\begin{equation}
{\rm ad} G_{{\cal S}}=\bigoplus_j \tau_j \otimes {\cal T}_j. 
\label{1.2.7}
\end{equation}
Since the light fermionic matter is given by the zero modes of the Dirac operator on 
${\cal S}$ it follows that the fermionic spectrum is given by 
\begin{eqnarray}
&&\bar\eta_{\dot\alpha, \tau_j}\in H^0({\cal S}, T_j), \nonumber\\
&&\psi_{\alpha, \tau_j} \in H^1({\cal S}, T_j), \nonumber\\
&&\bar\chi_{\dot\alpha, \tau_j} \in H^2 ({\cal S}, T_j), 
\label{1.2.8}
\end{eqnarray}
where $T_j$ is the vector bundle on ${\cal S}$ whose sections transform in the 
representation ${\cal T}_j$ of the structure group $H_{{\cal S}}$. The upper 
index in the cohomology 
groups $H^i$ is due to the fact that the fermions are twisted. 
Of course, the spectrum in eq.~\eqref{1.2.8} has to be supplemented by the complex conjugate
fields. Note that the term in eq.~\eqref{1.2.7} corresponding to $\tau_j={\bf 1}$, 
${\cal T}_j={\rm ad} H_{{\cal S}}$ 
counts the vector bundle moduli. As the result, the chiral spectrum 
is given by~\cite{BHV}
\begin{equation}
H^0 ({\cal S}, T_j^{\vee})^{\vee}
\oplus 
H^1 ({\cal S}, T_j) \oplus
H^2 ({\cal S}, T_j^{\vee})^{\vee}
\label{1.2.9}
\end{equation}
and the antichiral spectrum  
is given by~\cite{BHV}
\begin{equation}
H^0 ({\cal S}, T_j)
\oplus 
H^1 ({\cal S}, T_j^{\vee})^{\vee} \oplus
H^2 ({\cal S}, T_j),
\label{1.2.10}
\end{equation}
where the symbol $\vee$ stands for the dual bundle or vector space. 
The difference between the chiral and 
antichiral matter in the representation $\tau_j$ of $\Gamma_{{\cal S}}$ is
given by the difference of the Euler characteristics
\begin{equation}
n_{\tau_j}-n_{\tau_j^*}=\chi({\cal S}, {\cal T}_j^{\vee})-\chi({\cal S}, {\cal T}_j)=
-\int c_1({\cal T}_j) c_1 ({\cal S}), 
\label{1.2.11}
\end{equation}
where $c_1({\cal T}_j)$ and $c_1({\cal S})$ are the first Chern classes of ${\cal T}_j$
and the holomorphic tangent bundle of ${\cal S}$ respectively.

The analysis of the theory of the seven-branes wrapping ${\cal S}^{\prime}$ is identical to the one 
presented above.

Let us now discuss the theory localized on ${\mathbb R}^{1, 3}\times \Sigma$. It was 
also obtained in~\cite{BHV} by twisting the maximally supersymmetric gauge theory in six dimensions.
The result is that on the intersection  one gets two chiral multiplets 
\begin{equation}
(\sigma, \lambda_{\alpha}), \quad (\sigma^c, \lambda_{\alpha}^c), 
\label{1.2.12}
\end{equation}
where all these fields transform in representation of 
$G_{{\cal S}}\times G_{{\cal S}^{\prime}}$. An additional important feature is that 
the fields in~\eqref{1.2.12} are twisted and transform as sections of 
$K_{\Sigma}^{1/2}$, where $K_{\Sigma}$ is the canonical bundle on $\Sigma$. 
Note that since $\Sigma$ is a Riemann surface it is a spin manifold and
the square root of the canonical bundle exists. To obtain which representations 
of $G_{{\cal S}}\times G_{{\cal S}^{\prime}}$ are allowed one needs to know 
how the singularity is enhanced along $\Sigma$. Let the singularity be enhanced
to another $ADE$-type group $G_{\Sigma} \supset G_{{\cal S}} \times G_{{\cal S}^{\prime}}$. 
To obtain the allowed representations of $G_{{\cal S}} \times G_{{\cal S}^{\prime}}$ 
we decompose ${\rm ad} G_{\Sigma}$ as 
\begin{equation}
{\rm ad} G_{\Sigma} ={\rm ad} G_{{\cal S}}\oplus {\rm ad} G_{{\cal S}^{\prime}}\oplus
\bigoplus_j ({\cal U}_j \otimes {\cal U}_j^{\prime}). 
\label{1.2.13}
\end{equation}
The bifundamentals $(\sigma, \lambda_{\alpha})$, $(\sigma^c, \lambda^c_{\alpha})$
transform in the representations of $G_{{\cal S}} \times G_{{\cal S}^{\prime}}$
given by the non-adjoint summand 
$\bigoplus_j ({\cal U}_j \otimes {\cal U}_j^{\prime})$ in~\eqref{1.2.13}.
To obtain the low-energy field theory in four dimension we compactify $I_{\Sigma}$ on $\Sigma$. 
As was discussed above, we can put non-trivial instantons on both ${\cal S}$ and 
${\cal S}^{\prime}$ with structure groups $H_{{\cal S}}$ and $H_{{\cal S}^{\prime}}$
respectively. Then the matter fields originating from the intersection multiplets~\eqref{1.2.12}
will transform in representations of $\Gamma_{{\cal S}}\times \Gamma_{{\cal S}^{\prime}}$, 
where $\Gamma_{{\cal S}}$ is the commutant of $H_{{\cal S}}$ in $G_{{\cal S}}$ and 
$\Gamma_{{\cal S}^{\prime}}$ is the commutant of 
$H_{{\cal S}^{\prime}}$ in $G_{{\cal S}^{\prime}}$. More precisely, we decompose 
\begin{equation}
{\cal U} \otimes {\cal U}^{\prime} = \bigoplus_k (\nu_k, {\cal V}_k), 
\label{1.2.14}
\end{equation}
where $\nu_k$ is a representation of $\Gamma=
\Gamma_{{\cal S}}\times \Gamma_{{\cal S}^{\prime}}$ and ${\cal V}_k$ is a representation of 
$H=H_{{\cal S}} \times H_{{\cal S}^{\prime}}$. The chiral fermions in the representation
$\nu_k$ correspond to the zero modes of the Dirac operator on $\Sigma$. Thus, 
\begin{eqnarray}
&&\lambda_{\alpha, \nu_k} \in H^0
(\Sigma, K_{\Sigma}^{1/2}\otimes V_k), 
\label{1.2.15}\\
&&\lambda^c_{\alpha, \nu_k} \in H^0
(\Sigma, K_{\Sigma}^{1/2}\otimes V^{\vee}_k) \simeq 
H^1
(\Sigma, K_{\Sigma}^{1/2}\otimes V_k)^{\vee}, 
\label{1.2.16}
\end{eqnarray}
where in the last step in~\eqref{1.2.16} we have used the Serre duality on $\Sigma$
and $V_k$ is the vector bundle whose sections transform in the representation
${\cal V}_k$ of $H$. The additional factor $K_{\Sigma}^{1/2}$ in eqs.~\eqref{1.2.15} and~\eqref{1.2.16}
is due to the fact that the fermions are twisted. The difference between chiral and antichiral
matter in the representation $\nu_k$ of the low-energy gauge group $\Gamma$ is given by
the Euler characteristic
\begin{equation}
n_{\nu_k}-n_{\nu_k^*}=\chi(\Sigma, K_{\Sigma}^{1/2}\otimes V_k). 
\label{1.2.17}
\end{equation}
This concludes our review of the theory on the intersecting seven-branes. 
Additional details can be found in~\cite{BHV}. 


\section{Dynamical SUSY Breaking}


\subsection{Field Theory Requirements}


In this subsection, we will give a brief review of dynamical supersymmtery breaking 
in ${\cal N}=1$ SQCD following~\cite{ISS}. 
The goal is to formulate the field theory requirements which we will intend to realize 
on F-theory seven-branes. We consider ${\cal N}=1, SU(N_c)$ SQCD with $N_f$ 
fundamental flavors $Q, \tilde{Q}$ in the free magnetic range~\cite{Seiberg1, Seiberg2}
\begin{equation}
N_c+1 \leq N_f < \frac{3}{2}N_c.
\label{2.1.1}
\end{equation}
The flavors are taken to be massive and have a quadratic superpotential
\begin{equation}
W={\rm Tr} m M, 
\label{2.1.2}
\end{equation}
where 
\begin{equation}
M=Q_f \cdot \tilde Q_g, \quad f, q=1, \dots N_f. 
\label{2.1.3}
\end{equation}
This theory is known to have $N_c$ supersymmetric vacua with
\begin{equation}
\left< M\right>= \big(\Lambda^{3N_c-N_f} \det m \big)^{1/N_c}
m^{-1},
\label{2.1.4}
\end{equation}
where $\Lambda$ is the strong coupling scale. It was shown in~\cite{ISS}
that, in addition, this theory has a metastable SUSY breaking vacuum.
This was established by studying the
Seiberg dual~\cite{Seiberg1, Seiberg2} of the original
theory.  
The Seiberg dual theory is $SU(N_f-N_c)$ SQCD with $N_f$
fundamentals $q$, $\tilde{q}$ and $N_f^2$ extra singlets $\Phi_f^g$. It
has a quadratic leading order Kahler potential and the superpotential
given by (up to some field redefinition)
\begin{equation}
W_{{\rm dual}}=h {\rm Tr} q \Phi \tilde{q} -h \mu^2 {\rm Tr} \Phi,
\label{2.1.5}
\end{equation}
where $\mu=\sqrt{m\Lambda}$ and $h$ is a dimensionless parameter
(see~\cite{ISS} for additional details).
For simplicity, we have assumed that all eigenvalues of the mass
matrix are equal. This theory breaks supersymmetry by a rank condition
mechanism since F-flatness for $\Phi$ requires that
\begin{equation}
\tilde{q}^f q_g =\mu^2 \delta^f_g, 
\label{2.1.6}
\end{equation}
which cannot be satisfied because the number of colors of the dual
theory $N_f-N_c$ is less than the number of flavors $N_f$. However, it
was shown in~\cite{ISS} that there exists a metastable
SUSY breaking vacuum with
\begin{equation}
V_{min}=N_c \big| h^2\mu^4 \big|. 
\label{2.1.7}
\end{equation}
This result can be well-trusted in the regime
\begin{equation}
\epsilon \sim 
\sqrt{ \left| \frac{m}{\Lambda} \right| } \ll 1,
\label{2.1.8}
\end{equation}
These results were also generalized in~\cite{ISS} for
SQCD with gauge groups $SO(N_c)$ and $Sp(N_c)$. 
In this paper, we will concentrate
on the $SU(N_c)$ theories.


\subsection{Embedding in F-Theory Compactifications}


In the rest of the section, we will discuss how the field theory reviewed above can be 
obtained on the intersecting seven-branes in F-theory. We would like to build
a massive $SU(N_c)$ SQCD with fundamental matter so that the requirement~\eqref{2.1.1} is satisfied. 
In addition, the fundamental fields have to be very light. 
As we discussed in the previous section, the charged matter is in one-to-one correspondence 
with various bundle cohomology groups. The dimensions of bundle cohomology groups 
are not topological invariants. Thus, they depend on the location in the vector bundle and 
complex structure moduli spaces. As we move in the moduli space the dimensions might jump meaning 
that some extra charged matter fields might become light. Physically, this means that 
a certain number of matter fields have a quadratic superpotential with the mass matrix
depending on the moduli of the vector bundle on the complex structure
of the F-theory four-fold $X$. Somewhere in the moduli space, some eigenvalues of the mass
matrix can vanish increasing the number of the massless fields. If the difference between the 
chiral and anti-chiral fermions in some representation of the low-energy gauge group 
$\Gamma$ is non-zero then a certain number of matter fields will always stay massless
since they are protected by the topological invariants~\eqref{1.2.11} or~\eqref{1.2.17}. 
Hence, to build SQCD, we are interested in F-theory models with vanishing topological 
invariants~\eqref{1.2.11} and~\eqref{1.2.17}. Furthermore, we are interested in models 
where at a generic point in the moduli space all matter is massive. However, 
near some subvarieties the mass of some fields has to be become light which is also a requirement
to generate the field theory from the previous subsection.
Note that moduli are dynamical fields in the four-dimensional low-energy 
fields theory. However, eventually, we are interested in compactifications in which all the modul
are stabilized. Thus, we will assume that it is indeed the case and will view them as parameters. 
We will not discuss the issues of moduli stabilization in this paper. 

Let us point out that from a conceptual viewpoint engineering of SQCD with 
massive flavors in the context of F-theory is not much different from that 
in the case of flat non-compact branes studied 
in~\cite{OO1, Franco1, Bena, Ahn1, Ahn3, Giveon}. 
In both cases one has to take a certain number of intersecting branes and by using 
open string moduli engineer a mass term for the charged matter fields.
In the context of~\cite{OO1, Franco1, Bena, Ahn1, Ahn3, Giveon} the corresponding 
open string moduli are brane separations and in the context of F-theory 
the open string moduli are the vector bundle moduli. However, from a technical viewpoint 
our case is, clearly, more complicated. Since we are interested in quasi-realistic particle 
physics compactifications, the seven-branes have to wrap non-trivial compact four-cycles. 
In addition, these four-cycles are endowed with a non-trivial vector bundle. 
Since the mass term for the fundamental fields is controlled by the vector bundle 
moduli to understand the structure of 
the quadratic superpotential one has to take into a account 
the geometric properties of the four-cycle and of the vector bundle.

Below we will discuss a class of F-theory models which can lead to SQCD on the intersecting seven-branes
with requirements formulated in the previous subsection. In the rest of the section, 
we will give a general 
consideration. 
In the next section we will apply the ideas developed in this section 
for a concrete example of geometry of the seven-branes.


\subsection{The Spectrum Localized on the Surfaces}


First, we will consider the sector of the theory that comes from the surfaces
${\cal S}$ and ${\cal S}^{\prime}$. 
Let $V$ be 
an instanton on ${\cal S}$ with structure group $H_{{\cal S}}$ and $V^{\prime}$ 
be an instanton on ${\cal S}^{\prime}$ with structure group $H_{{\cal S}^{\prime}}$. 
If $G_{{\cal S}}$ and $G_{{\cal S}^{\prime}}$ are the singularities along 
${\cal S}$ and ${\cal S}^{\prime}$, the low-energy gauge group is 
$\Gamma_{{\cal S}} \times \Gamma_{{\cal S}^{\prime}}$ with 
$\Gamma_{{\cal S}} (\Gamma_{{\cal S}^{\prime}})$ being the commutant 
of $H_{{\cal S}}(H_{{\cal S}^{\prime}})$ in $G_{{\cal S}} (G_{{\cal S}^{\prime}})$. 
At this point let us simplify our model. 
For concreteness, let us choose the singularity 
along both ${\cal S}$ and ${\cal S}^{\prime}$ to be of the $A$-type. We will
assume that one of the factors in 
$\Gamma_{{\cal S}} \times \Gamma_{{\cal S}^{\prime}}$, 
say $\Gamma_{{\cal S}}$, is trivial 
as far as the gauge theory dynamics is concerned. There are several natural ways to achieve it.
The simplest way is to put an instanton on ${\cal S}$ with structure group $G_{{\cal S}}$. 
This way we find 
that $\Gamma_{{\cal S}}$ is completely broken. We also do not obtain any massless matter 
in the ${\cal S}$-sector except for the vector bundle moduli. One more way is to take 
$\Gamma_{{\cal S}}$ to be $U(1)$. Since $U(1)$ is infrared free it does not effect
the strong coupling dynamics of any non-Abelian factor and, hence, can be ignored. 
Another way is to assume that the volume of ${\cal S}$ is much bigger than the volume of 
${\cal S}^{\prime}$. Then the coupling constant of $\Gamma_{{\cal S}}$ is parametrically 
much smaller than the coupling constant of $\Gamma_{{\cal S}^{\prime}}$. 
In this case, $\Gamma_{{\cal S}}$ can be viewed as an (approximate) global symmetry. 
In this paper, we will concentrate on the first possibility. That is, we will put an instanton
on ${\cal S}$ with structure group $G_{{\cal S}}$ which we will denote $SU(n)$
\begin{equation}
H_{{\cal S}}=
G_{{\cal S}}=SU(n). 
\label{n0}
\end{equation}
In principle, one can keep the ${\cal S}^{\prime}$-sector 
non-trivial and generate SQCD with the product
gauge group which also might break SUSY in the 
infrared~\cite{OO, Ahn1, Ahn2, Franco}. However, we will simplify our model
and concentrate on the theory of~\cite{ISS} reviewed in the previous subsection. 

Furthermore, we will 
put the trivial instanton 
on the other surface ${\cal S}^{\prime}$.
Then the gauge group
$\Gamma_{{\cal S}^{\prime}}$ equals $G_{{\cal S}^{\prime}}$ which we will denote $SU(N_c)$. 
That is,
\begin{equation}
\Gamma_{{\cal S}^{\prime}}=G_{{\cal S}^{\prime}}=SU(N_c). 
\label{n1}
\end{equation}
Let us study the spectrum of the theory. In the ${\cal S}^{\prime}$-sector
we obtain ${\cal N}=1$, $SU(N_c)$ supersymmetric gauge theory without any matter. 
In the ${\cal S}$-sector the only 
fields are the moduli of $V$, which we view as parameters.

Let us now explain why we have chosen to put the trivial instanton on 
${\cal S}^{\prime}$. For this we have to see how the spectrum of the theory 
in the ${\cal S}^{\prime}$-sector gets modified if the instanton $V^{\prime}$ 
on ${\cal S}^{\prime}$ is non-trivial. Let us specify the low-energy
gauge group $\Gamma_{{\cal S}^{\prime}}$ to be $SU(N_c)$ as before. Since 
in the presence of a non-trivial vector bundle on ${\cal S}^{\prime}$
it does not coincide with $G_{{\cal S}^{\prime}}$, we will denote 
$G_{{\cal S}^{\prime}}=SU(N)$, $N>N_c$. The structure group of the vector 
bundle is then $SU(N-N_c)$. Note that, to be precise, the low-energy gauge group in this 
case is $SU(N_c)\times U(1)$ but as we discussed the $U(1)$-factor is irrelevant for our purposes 
and will be ignored. 
The spectrum of the theory in the ${\cal S}^{\prime}$-sector 
consists of the $SU(N_c)$ vector multiplet, the moduli of the vector bundle $V^{\prime}$
that we put on ${\cal S}^{\prime}$
and the matter fields charged under $SU(N_c)$.
According to our consideration 
in the previous section, in order to obtain the matter content,
we have to decompose the adjoint representation of $SU(N)$ 
under $SU(N_c)\times SU(N-N_c)$. The fields charged under $SU(N_c)$ are contained 
in the following terms of the decomposition (ignoring the $U(1)$-charges)
\begin{equation}
(\bf{N_c}, \bf{\overline{N-N_c}})\oplus (\bf{\overline{N}_c}, \bf{N-N_c}). 
\label{n2}
\end{equation}
Thus, the matter charged under 
$\Gamma_{{\cal S}^{\prime}}=SU(N_c)$ corresponds to the cohomology groups
with coefficients in $V^{\prime}$ and $V^{\prime \vee}$. 
From the previous section it follows that
the number of fields in the fundamental representation of $\Gamma_{{\cal S}}$ 
is~\footnote{Throughout the paper we denote by $H^i$ cohomology groups
and by $h^i$ their dimension.}
\begin{equation}
h^0 ({\cal S}^{\prime}, V^{\prime})+
h^1 ({\cal S}^{\prime}, V^{\prime \vee})+h^2 ({\cal S}^{\prime}, V^{\prime})
\label{n3}
\end{equation}
whereas the number of the antifundamental fields is 
\begin{equation}
h^0 ({\cal S}^{\prime}, V^{\prime \vee})+
h^1 ({\cal S}^{\prime}, V^{\prime})+h^2 ({\cal S}^{\prime}, V^{\prime \vee}).
\label{n4}
\end{equation}
Since $V^{\prime}$ is a stable bundle, 
$h^0({\cal S}^{\prime}, V^{\prime})=h^0({\cal S}^{\prime}, V^{\prime \vee})=0$. 
Furthermore, using the Serre duality we have 
\begin{equation}
h^2({\cal S}^{\prime}, V^{\prime \vee})=
h^0({\cal S}^{\prime}, V^{\prime} \otimes K_{{\cal S}^{\prime}}), 
\label{n5}
\end{equation}
where $K_{{\cal S}^{\prime}}$ is the canonical bundle on ${\cal S}^{\prime}$. In many cases
$V^{\prime} \otimes K_{{\cal S}^{\prime}}$ 
also does not have sections and the right hand side in~\eqref{n5} 
vanishes. For instance,
this is the case if ${\cal S}^{\prime}$ 
is one of del Pezzo surfaces~\cite{BHV}. 
We will assume that $h^2({\cal S}^{\prime}, V^{\prime})=
h^2({\cal S}^{\prime}, V^{\prime \vee})=0$. 
Then the matter charged under $\Gamma_{{\cal S}^{\prime}}=SU(N_c)$ receives a contribution 
only from $h^1 ({\cal S}^{\prime}, V^{\prime})$ 
and $h^1 ({\cal S}^{\prime}, V^{\prime \vee})$. It then follows that 
it is given by the Euler characteristics
\begin{equation}
h^1 ({\cal S}^{\prime}, V^{\prime})=-\chi({\cal S}^{\prime}, V^{\prime})
\label{n6}
\end{equation}
and 
\begin{equation}
h^1 ({\cal S}^{\prime}, V^{\prime \vee})=-\chi({\cal S}^{\prime}, V^{\prime \vee}).
\label{n7}
\end{equation}
In other words, the number of (anti)-fundamentals is given by topological invariants
and protected from becoming massive unless 
$\chi({\cal S}^{\prime}, V^{\prime})=\chi({\cal S}^{\prime}, V^{\prime \vee})=0$. 
This explains why we did not put a non-trivial instanton on 
the same seven-brane ${\cal S}^{\prime}$ which carries the $SU(N_c)$ gauge theory.
Putting a 
non-trivial vector bundle on ${\cal S}^{\prime}$ would generate (anti)-fundamental
matter of the gauge group $SU(N_c)$. This matter would be topologically protected 
from becoming massive unless 
$\chi({\cal S}^{\prime}, V^{\prime})=\chi({\cal S}^{\prime}, V^{\prime \vee})=0$ which is a 
strong restriction. Hence, it would be  
difficult to generate SQCD with massive flavors in this case. 

To summarize, the spectrum of the theory localized on the surfaces is pure 
$SU(N_c)$ gauge theory with some number of vector bundle moduli. 
The fundamental matter charged under $SU(N_c)$ comes from the sector localized on the intersection 
curve $\Sigma$ which we have to discuss next. 


\subsection{The Spectrum Localized on the Curve}


Now let us discuss the theory in the $\Sigma$-sector. 
First, we need to specify the 
enhanced singularity along $\Sigma$. We chose the singularities along ${\cal S}$,
${\cal S}^{\prime}$ and $\Sigma$ to be of the $A$-type. 
In notation of the previous subsection,
the matter fields on $\Sigma$, which we denote as 
$(\sigma, \lambda_{\alpha})$ and $(\sigma^c, \lambda^c_{\alpha})$, transform as the 
(anti)-fundamentals 
of the group $SU(n) \times SU(N_c)$. 
When we compactify to four dimensions, the $SU(N_c)$ factor survives as the gauge symmetry and
the $SU(n)$ factor is completely broken by the vector bundle.
Therefore,
the massless four-dimensional fields transform as
(anti)-fundamentals of $SU(N_c)$.
Indeed, the non-adjoint summand in eq.~\eqref{1.2.13} 
is our case is
\begin{equation}
({\bf N_c}, {\bf{\overline{n}}})\oplus ({\bf \overline{N}_c}, \bf{n})
\label{2.2.0}
\end{equation}
The fields 
corresponding to the first terms are $(\sigma, \lambda_{\alpha})$ 
and the fields corresponding to the second term are  $(\sigma^c, \lambda^c_{\alpha})$.
When we compactify on $\Sigma$,
the first entry in both terms in~\eqref{2.2.0} labels the representation 
of the low-energy gauge group $SU(N_c)$ and the second entry 
specifies the vector bundle.
Recalling that the fields on the intersection 
are twisted by the square root of the canonical bundle,
we obtain the following matter content.
We have the multiplets $(\tilde{Q}, \tilde{\lambda}_{\alpha})$ 
whose number is determined by 
$h^0 (\Sigma, K^{1/2}_{\Sigma}\otimes V|_{\Sigma})$ and the multiplets 
$(Q, \lambda_{\alpha})$ whose number is determined by 
$h^1 (\Sigma, K^{1/2}_{\Sigma}\otimes V|_{\Sigma})$ (or $h^0 
(\Sigma, K^{1/2}_{\Sigma}\otimes V^{\vee}|_{\Sigma})$).
Here $V|_{\Sigma}$ is $V$ restricted to $\Sigma$. The fields  $(Q, \lambda_{\alpha})$
transform in the fundamental representation of $SU(N_c)$ and the fields 
$(\tilde{Q}, \tilde{\lambda_{\alpha}})$ transform in the antifundamental representation of 
$SU(N_c)$. To generate SQCD, we need the number of fundamental and antifundamental 
multiplets to be the same. This means that the Euler characteristic
\begin{equation}
\chi(\Sigma, K^{1/2}_{\Sigma}\otimes V|_{\Sigma})=
h^0 (\Sigma, K^{1/2}_{\Sigma}\otimes V|_{\Sigma})-
h^1 (\Sigma, K^{1/2}_{\Sigma}\otimes V|_{\Sigma})
\label{2.2.2}
\end{equation}
has to vanish. From the Riemann-Roch theorem (see, for example,~\cite{GH}) it follows that
\begin{eqnarray}
\chi(\Sigma, K^{1/2}_{\Sigma}\otimes V|_{\Sigma})& = &
(1-g) c_0 (K_{\Sigma}^{1/2} \otimes V|_{\Sigma}) +
 c_1 (K_{\Sigma}^{1/2} \otimes V|_{\Sigma}) \nonumber \\
 & = & (1-g)c_0(V|_{\Sigma})+c_1 (V|_{\Sigma})+ c_0(V|_{\Sigma})c_1(K_{\Sigma}^{1/2}), 
\label{2.2.3}
\end{eqnarray}
where $g$ is the genus of $\Sigma$. In this paper, we will consider the case 
\begin{equation}
\Sigma \simeq {\mathbb P}^1. 
\label{2.2.3.1}
\end{equation}
Then, taking into account that
\begin{equation}
K_{\Sigma}^{1/2}\simeq {\cal O}(-1)
\label{2.2.4}
\end{equation}
it follows from~\eqref{2.2.3}
that $\chi(\Sigma, K^{1/2}_{\Sigma}\otimes V|_{\Sigma})=0$ if
\begin{equation}
c_1(V|_{\Sigma})=0.
\label{2.2.4.1}
\end{equation}
This condition is trivially satisfied if $V$ has structure group $SU(n)$. 

Thus, the question of analyzing how to obtain SQCD with appropriate number 
of light fields is reduced to analyzing under what conditions
the vector bundle
\begin{equation}
{\cal O}(-1) \otimes V|_{\Sigma}
\label{2.2.5}
\end{equation}
has global holomorphic sections.
This problem is known to arise in a different context, namely,
under what conditions the non-perturbative superpotential due to a string 
(open membrane) instanton in heterotic M-theory does or does not 
vanish~\cite{Greene, Witten99, BDO2, BDO3}. In that context, $\Sigma$ is an isolated
sphere inside the Calabi-Yau threefold on which the $E_8 \times E_8$
heterotic string theory is compactified, 
$V$ 
is a vector bundle on the threefold and 
$h^0(\Sigma, {\cal O}(-1) \otimes V|_{\Sigma})$ counts the number of the zero modes 
of the Dirac operator coupled to the world-sheet fermions on $\Sigma$. 
The existence or non-existence of the global sections of $ {\cal O}(-1) \otimes V|_{\Sigma}$
depends on the moduli of $V$ and on the complex structure of the Calabi-Yau threefold. 
This problem was analyzed in detail for some geometries in~\cite{BDO2, BDO3} 
where the dependence of the non-perturbative superpotential on the vector bundle 
moduli was explicitly calculated.
In the next section we will specify the details of our examples 
so that the set-up is reduced to the one studied in~\cite{BDO2, BDO3}. This will allow us 
to calculate the locus in the moduli space near which the right number 
of the fundamental fields becomes light, realizing this way SQCD in the free magnetic range. 

To finish this section, let us discuss the quadratic superpotential for the 
(anti)-fundamental fields $Q$ and $\tilde{Q}$. 
As was shown in~\cite{BHV}, the
action $I_{\Sigma}$ contains terms which give rise to the superpotential. 
This superpotential can be written as follows. Let $\omega_Q$ be the element
of $H^0(\Sigma, K_{\Sigma}^{1/2} \otimes V^{\vee}|_{\Sigma})$ corresponding to $Q$. That is, the
world-volume field $\sigma$ is written as
\begin{equation}
\sigma= \sum Q \cdot \omega_Q, 
\label{2,2.6}
\end{equation}
where the sum is over all zero modes of $\sigma$. Similarly, let 
$\omega_{\tilde{Q}}$ be the element in 
$H^0(\Sigma, K_{\Sigma}^{1/2} \otimes V|_{\Sigma})$ corresponding to 
$\tilde{Q}$. 
Note that
\begin{equation}
\omega_Q \cdot \omega_{\tilde{Q}} \in H^0 (\Sigma, K_{\Sigma}\otimes
(V \otimes V^{\vee})|_{\Sigma})
\label{2.2.8}
\end{equation}
At last, let $\omega_{\phi}$ be the differential form corresponding to the 
vector bundle modulus $\phi$.
It is a standard result that~\footnote{See, for example, section 15.7.3 of~\cite{GSW}.} 
\begin{equation}
\omega_{\phi} \in H^1({\cal S}, End V)=
H^1 ({\cal S}, V \otimes V^{\vee})
\label{2.2.9}
\end{equation}
If we restrict eq.~\eqref{2.2.9} to $\Sigma$ and use the Serre dulity
\begin{equation}
H^1 (\Sigma, (V \otimes V^{\vee})|_{\Sigma})\simeq 
H^0 (\Sigma, K_{\Sigma}^{1/2}\otimes (V \otimes V^{\vee})|_{\Sigma})^{\vee}
\label{2.2.9.1}
\end{equation}
we that one can pair up elements in~\eqref{2.2.8} and~\eqref{2.2.9.1} 
to obtain a complex number since they parametrize the spaces dual to each other. 
That is, we have the following natural map 
\begin{equation}
H^0(\Sigma, K_{\Sigma}^{1/2} \otimes V|_{\Sigma}) 
\otimes 
H^1(\Sigma, (V \otimes V^{\vee})|_{\Sigma})
\otimes
H^0(\Sigma, K_{\Sigma}^{1/2} \otimes V^{\vee}|_{\Sigma}) 
\to {\mathbb C}. 
\label{2.2.80}
\end{equation}
Explicitly, it can be done as follows. We have
\begin{equation}
\omega_Q \cdot \omega_{\tilde{Q}}\in 
H^0 (\Sigma, K_{\Sigma}\otimes
(V \otimes V^{\vee})|_{\Sigma})\simeq H^{(1, 0)}_{\bar \partial}
(\Sigma, (V \otimes V^{\vee})|_{\Sigma})
\label{2.2.10}
\end{equation}
Hence, we can view $\omega_Q \cdot \omega_{\tilde{Q}}$ 
as a $(1, 0)$ differential form on $\Sigma$. On the other hand, 
\begin{equation}
\omega_{\phi} \in H^1(\Sigma, (V \otimes V^{\vee})|_{\Sigma})\simeq 
 H^{(0, 1)}_{\bar \partial}
(\Sigma, (V \otimes V^{\vee})|_{\Sigma}).
\label{2.2.11}
\end{equation}
Hence, we can view $\omega_{\phi}$ as a $(0, 1)$-differential form on $\Sigma$. 
Thus, the superpotential can be written as 
\begin{equation}
W=\lambda \phi {\rm Tr} (Q \cdot \tilde{Q}), 
\label{2.2.12}
\end{equation}
where
\begin{equation}
\lambda=\int_{\Sigma} \omega_Q \cdot \omega_{\tilde{Q}} \wedge \omega_{\phi}, 
\label{2.2.13}
\end{equation}
where we suppressed the flavor indices. 


\subsection{The Summary of the Model}


In this small subsection, we will summarize the ingredients necessary to build SQCD 
found above. The various pieces of the spectrum come from the three different sources, 
the surface ${\cal S}^{\prime}$, the surface ${\cal S}$ and the intersection curve $\Sigma$. 
More precisely the role of each of them is as follows. 
\begin{itemize}
\item The surface ${\cal S}^{\prime}$ contributes ${\cal N}=1$, 
$SU(N_c)$ gauge theory. 
\item The surface ${\cal S}$ contributes vector bundle moduli. 
\item The intersection curve $\Sigma$ contributes matter fields $Q$ and $\tilde{Q}$ in the fundamental 
and antifundamental representations of the gauge group $SU(N_c)$. The mass of these fields 
is controlled by the vector bundle moduli through the superpotential~\eqref{2.2.12}. 
\end{itemize}
%


\section{An F-Theory Realization of SQCD in the Free Magnetic Range}


\subsection{The Geometric Data}


In this section, we will give a realization of the ideas developed in the 
previous section. 
From the above consideration it follows that the 
details of the geometry of the surface ${\cal S}^{\prime}$ 
are irrelevant. The role of it is to produce the $SU(N_c)$ vector multiplet. 
Therefore, we only need to specify
the surface ${\cal S}$ and the 
curve $\Sigma \subset {\cal S}$. Motivated by the heterotic-F-theory
duality it is reasonable to choose ${\cal S}$ to be the base of an elliptically fibered 
Calabi-Yau threefold. We will choose ${\cal S}$ to be rational elliptic surface $d P_9$ which 
is known to a be a possible base for a Calabi-Yau threefold. 
Various properties of $d P_9$ can be found, for example, in~\cite{Barth}.
It is worth pointing out that elliptically fibered over $d P_9$ 
Calabi-Yau threefolds as well as their quotient over a discrete group are often used
in GUT and Standard Model heterotic compactifications. In particular, 
such manifolds were used in constructing a heterotic standard model 
in~\cite{Braun1, Braun2, Braun3, Braun4, Ron1, Ron2}.

Let us present here some facts about $d P_9$.
The surface $dP_9$ is obtained
from ${\mathbb P}^2$ by blowing up nine distinct points. Thus, the basis of effective 
curves in $d P_9$ can be chosen to be 
\begin{equation}
\{\ell, e_1, \dots e_9 \},
\label{3.1.1}
\end{equation}
where $\ell$ is the hyperplane divisor inherited from ${\mathbb P}^2$ and $e_1, \dots e_9$ 
are the nine exceptional divisors each isomorphic to ${\mathbb P}^1$. However, it is more 
convenient to work with a different basis. The surface $d P_9$ admits an elliptic fibration 
over ${\mathbb P}^1$. We identify the base of this fibration, $\sigma$, with one 
of the exceptional curves, say $e_1$. Let $\pi$ be the projection map 
\begin{equation}
\pi: d P_9 \to \sigma =e_1.
\label{3.1.1.1}
\end{equation}
A more convenient basis is 
\begin{equation}
\{F, e_1, \dots e_9 \},
\label{3.1.4}
\end{equation}
where $F$ is the class of the elliptic fiber. In terms of the curves in the basis~\eqref{3.1.1}
it is given by 
\begin{equation}
F=3 \ell - \sum_{i=1}^9 e_i. 
\label{3.1.3}
\end{equation}
The intersection numbers of the curves in~\eqref{3.1.4} are given by
\begin{equation}
e_i \cdot e_j =-\delta_{i j}, \quad e_i \cdot F=1. 
\label{3.1.5}
\end{equation}
The Chern classes of $d P_9$ are given by 
\begin{equation}
c_1 (d P_9)=F, \quad c_2 (d P_9)=12. 
\label{3.1.2}
\end{equation}
Being an elliptic fibration, $d P_9$ can be described by the Weierstrass equation
\begin{equation}
y^2 z= x^3+ f x z^2 + g z^3, 
\label{3.1.5.1}
\end{equation}
where $f$ and $g$ are polynomials on the base $\sigma \simeq {\mathbb P}^1$. 
More precisely, $f$ is a polynomial of degree four and $g$ is a polynomial of degree six. 
Furthermore, $z, x$ and $y$ are sections of the following line bundles~\cite{FMW}
\begin{equation}
z \in H^0(d P_9, {\cal O}_{d P_9}(3 \sigma)), \quad 
x \in H^0(d P_9, {\cal O}_{d P_9}(3 \sigma+2 F)), \quad 
y \in H^0(d P_9, {\cal O}_{d P_9}(3 \sigma+3 F)). 
\label{3.1.5.2}
\end{equation}
One can see that each term in eq~\eqref{3.1.5.1} is a section of the same line bundle. 

After having specified the surface ${\cal S}$, we need to specify a genus zero curve 
$\Sigma \in {\cal S}$. We will choose it to be the base of $d P_9$, $\sigma$. That is, 
\begin{equation}
\Sigma=\sigma \simeq {\mathbb P}^1. 
\label{3.1.5.3}
\end{equation}

The next ingredient which needs to be specified
is an $SU(n)$ instanton $V$ on $d P_9$. Since $d P_9$ 
is elliptically fibered we can use the spectral 
cover construction~\cite{FMW, Donagi}. 
According to this construction, an $SU(n)$ vector bundle on elliptic 
$d P_9$ (or any elliptically fibered manifold) can be obtained from the spectral data 
\begin{equation}
({\cal C}, {\cal N}), 
\label{3.1.6}
\end{equation}
where the spectral cover ${\cal C}$ is an $n$-fold cover of the base $\sigma$
(in our case ${\cal C}$ is a Riemann surface)
and ${\cal N}$ is a line bundle on ${\cal C}$. The corresponding 
$SU(n)$ vector bundle $V$ can be obtained from the spectral data~\eqref{3.1.6}
by a Fourier-Mukai transformation~\cite{FMW, Donagi}.
We will choose the homology class of the spectral cover to be of the form
\begin{equation}
{\cal C}=n \sigma + k F. 
\label{3.1.7}
\end{equation}
The coefficient $n$ determines the rank of the vector bundle $V$ and
the coefficient $k$ can be shown to be the second Chern class
(the instanton number) of $V$. In order to make sure that the bundle $V$ 
is stable (that is, admits a connection solving the BPS equations~\eqref{1.2.6}) 
the homology class of the spectral cover has to contain irreducible curves. 
One can show that this is the case if the following condition is satisfied
\begin{equation}
k \geq n. 
\label{3.1.8}
\end{equation}
Let us work out some properties of ${\cal C}$. 
As we mentioned before, 
${\cal C}$ is simply a Riemann surface. For future reference, let us 
calculate its genus $g_{{\cal C}}$. It can be obtained using the 
adjunction and Riemann-Hurwitz formulas (see for example~\cite{GH}). 
From the adjunction formula it follows that the canonical bundle of ${\cal C}$ is 
\begin{equation}
K_{{\cal C}}=\big(K_{d P_9} \otimes {\cal O}_{d P_9} ({\cal C})\big)|_{{\cal C}}. 
\label{3.1.9}
\end{equation}
Therefore the degree of the canonical bundle of ${\cal C}$ is given by
\begin{equation}
{\rm deg} K_{{\cal C}}=K_{d P_9} \cdot {\cal C}+ {\cal C} \cdot {\cal C}. 
\label{3.1.10}
\end{equation}
Knowing the degree of the canonical bundle we can obtain the genus 
by the Riemann-Hurwitz formula 
\begin{equation}
2 g_{{\cal C}}-2 ={\rm deg} K_{{\cal C}}. 
\label{3.1.11}
\end{equation}
The from eqs.~\eqref{3.1.5}, \eqref{3.1.2}, \eqref{3.1.7}, \eqref{3.1.10} and~\eqref{3.1.11}
it follows that 
\begin{equation}
g_{{\cal C}}=n k -\frac{(n-1)(n+2)}{2}. 
\label{3.1.12}
\end{equation}
Now we will calculate how many parameters the linear system of ${\cal C}$ has. 
The number of {\it projective} parameters of the spectral cover is given by 
\begin{equation}
h^0 (d P_9, {\cal O}_{d P_9}({\cal C})). 
\label{3.1.13}
\end{equation}
This number can be calculated using a simple Leray spectral sequence according to which 
\begin{equation}
h^0 (d P_9, {\cal O}_{d P_9}({\cal C}))=h^0 (\sigma, \pi_* {\cal O}_{d P_9}({\cal C}))=
h^0 (\sigma, \pi_* {\cal O}_{d P_9}(n \sigma+k F)). 
\label{3.1.14}
\end{equation}
The direct image $\pi_* {\cal O}_{d P_9}(n \sigma+k F)$ was computed in Appendix C 
of~\cite{BDO3} by induction in $n$. Here we just quote the result
\begin{equation}
\pi_* {\cal O}_{d P_9}(n \sigma+k F)=
{\cal O}(k) \oplus \bigoplus_{i=2}^{n} {\cal O}(k-i). 
\label{3.1.15}
\end{equation}
Note that for $k \geq n$ all entries in the right hand side of~\eqref{3.1.15} are non-negative. 
Since $h^0 ({\mathbb P}^1, {\cal O}(i))=i+1$ for $i \geq 0$, it follows that 
\begin{equation}
h^0 (d P_9, {\cal O}_{d P_9}({\cal C}))=(k+1)+(k-1) + \dots + (k-n+1)=n k -\frac{(n+1)(n-2)}{2}. 
\label{3.1.16}
\end{equation}
The parameters of the spectral cover form a projective space 
${\mathbb P}^{h^0 (d P_9, {\cal O}_{d P_9}({\cal C}))-1}$~\cite{BDO1, BOR}. 
Later, we will introduce an explicit coordinate parametrization of this space. 

Now we move on to discussing the line bundle ${\cal N}$. 
An arbitrary choice of ${\cal N}$ on the spectral cover ${\cal C}$ will 
lead to a vector bundle $V$ on $d P_9$
with structure group $U(n)$. The condition 
under which ${\cal N}$ produces an $SU(n)$ rather than $U(n)$ vector bundle 
was derived in~\cite{FMW}. It can be formulated as follows: the degree of 
${\cal N}$ has to be related to the genus of the spectral cover as follows
\begin{equation}
{\rm deg} {\cal N}=g_{{\cal C}}-1+n. 
\label{3.1.18}
\end{equation}
The moduli space of line bundles ${\cal N}$ is the Jacobian 
${\cal J}_{g_{{\cal C}}} \simeq \big( {\mathbb T}^2 \big)^{g_{{\cal C}}}$.
Thus, the moduli space of the vector bundle $V$ is a Jacobian bundle over 
the projective space 
${\mathbb P}^{h^0 (d P_9, {\cal O}_{d P_9}({\cal C}))-1}$.
Unfortunately, it is very difficult to introduce an explicit parametrization of the Jacobian 
and have an analytic control over it. Therefore, at this step, we will simplify our analysis. 
We will fix the moduli of ${\cal N}$ at some particular values and study how 
$h^0(\sigma, K_{\sigma}^{1/2} \otimes V|_{\sigma})$ behaves only as we move in the projective 
space of the parameters of the spectral cover. We will fix the line bundles ${\cal N}$ on ${\cal C}$
as follows. We will choose ${\cal N}$ to be the restriction of the following discrete 
line bundles on $d P_9$. 
\begin{equation}
{\cal N}={\cal O}_{d P_9} \big( n(\frac{1}{2}+\lambda) \sigma+ 
(\frac{1}{2}-\lambda)k F +(\frac{1}{2}+ n \lambda)F\big), 
\label{3.1.19}
\end{equation}
where the discrete parameter $\lambda$ has to be chosen in such a way that the right hand 
side in~\eqref{3.1.19} is an integral class on $d P_9$. For example, if $n$ is odd
one always gets an integral class if $\lambda$ is half-integer. Starting this point, 
we will always mean by ${\cal N}$ a line bundle on $d P_9$ of the form~\eqref{3.1.19}
and the corresponding spectral line bundle on ${\cal C}$ we will denote as
${\cal N}|_{{\cal C}}$. It is straightforward to check using 
eqs.~\eqref{3.1.7}, \eqref{3.1.5} and~\eqref{3.1.9} that the degree of 
${\cal N}|_{{\cal C}}$ is indeed given by eq.~\eqref{3.1.18} independent of $\lambda$. 

To summarize, we will consider vector bundles $V$ on $d P_9$ constructed 
using the spectral data $({\cal C}$, ${\cal N}|_{{\cal C}})$, where 
${\cal C}$ is given by eq.~\eqref{3.1.7} and ${\cal N}|_{{\cal C}}$
is obtained by restriction of~\eqref{3.1.19} to ${\cal C}$.


\subsection{The Matter Localized on the Curve}


In this subsection, we will consider the matter localized on the curve $\sigma$. 
As was discussed before, 
it is determined by the cohomology groups
\begin{equation}
H^0 (\sigma, {\cal O}(-1) \otimes V|_{\sigma}), \quad H^1 (\sigma, {\cal O}(-1) \otimes V|_{\sigma}),
\label{3.3.1}
\end{equation}
where we have used the fact that $K_{\sigma}^{1/2}={\cal O}(-1)$. The analysis in 
this subsection will be similar to the one in~\cite{BDO2, BDO3} though the context is different. 
Our goal is to derive the equation in the moduli space of ${\cal C}$ along the zero locus 
of which one gets massless fundamental fields whereas away from this locus all the
fundamental fields are massive. 

The bundle $V|_{\sigma}$ can be obtained from the spectral data 
$({\cal C}, {\cal N}|_{{\cal C}})$ as follows~\cite{BDO2, BDO3}
\begin{equation}
V|_{\sigma}=\pi_{{\cal C}} {\cal N}|_{{\cal C}}, 
\label{3.3.2}
\end{equation}
where $\pi_{{\cal C}}:{\cal C} \to \sigma$ is the $n$-fold cover map. 
Then from a Leray spectral sequence it follows that
\begin{equation}
h^0(\sigma, {\cal O}(-1)\otimes V|_{\sigma})=
h^0 ({\cal C}, ({\cal N}\otimes {\cal O}_{d P_9}(-F))|_{{\cal C}}). 
\label{3.3.3}
\end{equation}
Denote 
\begin{equation}
{\cal N}(-F)={\cal N}\otimes {\cal O}_{d P_9}(-F). 
\label{3.3.4}
\end{equation}
Thus, we have to study under what conditions $h^0({\cal C}, {\cal N}(-F)|_{{\cal C}})$
vanishes. Note that the Euler characteristic of
$ {\cal N}(-F)|_{{\cal C}}$ vanishes. Indeed, from the Riemann-Roch formula 
\begin{equation}
\chi({\cal C}, {\cal N}(-F)|_{{\cal C}})=d-g_{{\cal C}}+1, 
\label{3.3.5}
\end{equation}
where by $d$ we denoted the degree of the line bundle ${\cal N}(-F)|_{{\cal C}}$.
Since the degree of ${\cal N}|_{{\cal C}}$ is $g_{{\cal C}}-1+n$ it follows that 
\begin{equation}
d=g_{{\cal C}}-1
\label{3.3.6}
\end{equation}
and, hence, the Euler characteristic in~\eqref{3.3.4} vanishes. 

The dimension $h^0({\cal C}, {\cal N}(-F)|_{{\cal C}})$
depends on the parameters of ${\cal C}$. As we move in the projective space
of these parameters, $h^0({\cal C}, {\cal N}(-F)|_{{\cal C}})$ might jump. 
We are interested in examples where $h^0(({\cal C}, {\cal N}(-F)|_{{\cal C}})$
is zero at a generic point in the moduli space and jumps along some subvariety. 
Let us now show how to derive the equation of this subvariety. First, we
will give some general discussion and then give a specific example. 

Consider the following short exact sequence on $d P_9$
\begin{equation}
0 \rightarrow E \otimes {\cal O}_{d P_9}(-D) \stackrel{f_D}{\rightarrow}
E \stackrel{r}{\rightarrow} E|_{D}\to 0. 
\label{3.3.7}
\end{equation}
Here $E$ is an arbitrary holomorphic vector bundle on $d P_9$ and $D$ is a
divisor in it. The map $r$ is just the restriction map. The map $f_D$ is a multiplication 
by a section of ${\cal O}_{d P_9}(D)$ which vanishes precisely on $D$. 
This sequence can be understood as follows. Let $e$ be any section of $E$. 
Let us restrict $e$ to $D$ and find the kernel of the restriction map. 
The kernel consists of such sections $e$ which vanish on $D$. Such sections can be written as
$e=f_D e^{\prime}$ for some $e^{\prime}$. It is clear that $e^{\prime}$ transforms 
with transition functions of $E \otimes {\cal O}_{d P_9}(-D)$. This means
that the kernel of $r$ is $E \otimes {\cal O}_{d P_9}(-D)$. For our purposes, 
we choose
\begin{equation}
E={\cal N}(-F), \quad D={\cal C}. 
\label{3.3.8}
\end{equation}
The sequence~\eqref{3.3.7} becomes
\begin{equation}
0 \rightarrow {\cal N}(-F-{\cal C}) \stackrel{f_{{\cal C}}}{\rightarrow}
{\cal N}(-F) \stackrel{r}{\rightarrow} {\cal N}(-F)|_{{\cal C}}\to 0, 
\label{3.3.9}
\end{equation}
where by $ {\cal N}(-F-{\cal C})$ we simply denoted $ {\cal N}(-F) \otimes {\cal O}_{d P_9}(-{\cal C})$.
From here we obtain the corresponding long exact sequence of the cohomology groups
\begin{eqnarray}
0 & \rightarrow & H^0(d P_9, {\cal N}(-F-{\cal C})) \rightarrow
H^0(d P_9, {\cal N}(-F)) \rightarrow 
H^0 ({\cal C}, {\cal N}(-F)|_{{\cal C}}) \nonumber \\
 & \rightarrow & H^1(d P_9, {\cal N}(-F-{\cal C})) \rightarrow
H^1(d P_9, {\cal N}(-F)) \rightarrow 
H^1 ({\cal C}, {\cal N}(-F)|_{{\cal C}})  \rightarrow \dots .
\label{3.3.10}
\end{eqnarray}
Note that the cohomology group $H^0 ({\cal C}, {\cal N}(-F)|_{{\cal C}})$ 
is exactly the object we are interested in. Also note that if 
$h^0 (d P_9, {\cal N}(-F))$ is non-zero, $h^0 ({\cal C}, {\cal N}(-F)|_{{\cal C}})$
cannot vanish. Hence, in this case we always have massless fundamental matter. 
Therefore, we will study the case when $h^0 (d P_9, {\cal N}(-F))=0$. Then the 
sequence~\eqref{3.3.10} simplifies and becomes
\begin{equation}
0 \rightarrow H^0 ({\cal C}, {\cal N}(-F)|_{{\cal C}}) \rightarrow
W_1 \stackrel{f_{{\cal C}}}{\rightarrow} W_2 \rightarrow \dots, 
\label{3.3.11}
\end{equation}
where $W_1$ and $W_2$ are the following vector spaces
\begin{equation}
W_1 =H^1(d P_9, {\cal N}(-F -{\cal C}))
\label{3.3.12.1}
\end{equation}
and 
\begin{equation}
W_2 =H^1(d P_9, {\cal N}(-F)).
\label{3.3.12.2}
\end{equation}
Both $W_1$ and $W_2$ are finite-dimensional vector spaces. The map $f_{{\cal C}}$ in~\eqref{3.3.11}
between them is a multiplication by a section of ${\cal O}_{d P_9}({\cal C)}$. 
It depends on the parameters of ${\cal C}$. This map can be organized as a finite-dimensional matrix. 
Thus, $h^0 ({\cal C},  {\cal N}(-F)|_{{\cal C}})$ is non-zero if the matrix 
$f_{{\cal C}}$ has a non-trivial kernel. We will consider the case when 
${\rm dim} W_1 = {\rm dim} W_2$. Then $f_{{\cal C}}$ is a square matrix. Therefore, 
$h^0 ({\cal C},  {\cal N}(-F)|_{{\cal C}})$ is non-zero if and only if 
\begin{equation}
{\rm det} f_{{\cal C}}=0. 
\label{3.3.13}
\end{equation}
Away from the locus given by eq.~\eqref{3.3.13} all fundamental fields $Q$ and $\tilde{Q}$ 
are very massive and the theory on the intersecting seven-branes is just pure 
$SU(N_c)$ supersymmetric Yang-Mills theory. Near the locus~\eqref{3.3.13} some number of the 
fundamental matter fields becomes light and the theory is SQCD 
with massive matter. This equation alone does not tell us exactly how many 
fundamental fields we obtain. We will discuss it later in this section. 
Now we will present an example of computation of ${\rm det} f_{{\cal C}}$~\cite{BDO2, BDO3}.

{\bf Example}

In this example, we will choose a vector bundle $V$ to have the structure group $SU(3)$. 
We will specify the second Chern class of $V$ to be $k=5$. In addition, we choose the discrete 
parameter $\lambda$ in~\eqref{3.1.19} to be $\lambda =\frac{3}{2}$. 
Then we obtain
\begin{eqnarray}
&&{\cal C}=3 \sigma + 5 F, \nonumber\\
&&{\cal N}(-F)={\cal O}_{d P_9} (6 \sigma -F),\nonumber \\
&&{\cal N}(-F-{\cal C})={\cal O}_{d P_9} (3 \sigma -6 F). 
\label{3.3.14}
\end{eqnarray}
Let us start with the explicit parametrization of the spectral cover. 
The number of the {\it projective} parameters is given by eq.~\eqref{3.1.16}. 
In our case it is $13$. Since from eq.~\eqref{3.1.15} we have
\begin{equation}
\pi_* {\cal O}_{d P_9} (3 \sigma +5 F)={\cal O}(5) \oplus {\cal O}(3) \oplus {\cal O}(2),
\label{3.3.15}
\end{equation}
we can write the equation for ${\cal C}$ as follows
\begin{equation}
{\cal C}=a_5 z+ a_3 x + a_2 y, 
\label{3.3.16}
\end{equation}
where $z$, $x$, $y$ are the variables in the Weierstrass equation~\eqref{3.1.5.1}, \eqref{3.1.5.2}. 
The coefficients $a_k$ are $a_k =\pi^* A_k$, where $A_k$ is a section of 
$H^0(\sigma, {\cal O}(k))$, 
that is 
a polynomial of degree $k$ on $\sigma \simeq {\mathbb P}^1$. Thus, 
if $(u, v)$ are projective coordinates on $\sigma$ then we have the following explicit 
parametrization of $A_k$ 
\begin{eqnarray}
&&A_5 = \psi_1 u^5 + \psi_2 u^4 v + \psi_3 u^3 v^2+ \psi_4 u^2 v^3+\psi_5   u v^4 +\psi_6 v^5, 
\nonumber \\
&&A_3 = \phi_1 u^3 + \phi_2 u^2 v + \phi_3 u v^2+ \phi_4 v^3, 
\nonumber \\
&&A_2 = \chi_1 u^2 + \chi_2 u v + \chi_3 v^3, 
\label{3.3.17}
\end{eqnarray}
where $\{\psi_a, \phi_b, \chi_c\}$ are the $13$ {\it projective} 
parameters of the spectral cover. The actual equation of ${\cal C}$ in $d P_9$
is obtained by 
setting~\eqref{3.3.16} to zero. This equation is invariant under rescaling of all 
$\{\psi_a, \phi_b, \chi_c\}$ by a non-zero complex number. Therefore, only $12$ parameters are 
independent. They parametrize the projective space ${\mathbb P}^{12}$. 

In the next step, we need to parametrize the vector spaces $W_1$ and $W_2$. 
The idea is to push $W_1$ and $W_2$ down to the base $\sigma\simeq {\mathbb P}^1$ 
where one can use a paramerization in terms of polynomials. From a Leray spectral sequence 
it follows that 
\begin{equation}
W_1 =H^1(d P_9, N(-F-{\cal C}))\simeq H^1 (\sigma, \pi_* {\cal N}(-F -{\cal C})). 
\label{3.3.18}
\end{equation}
To obtain this result we used the fact that $R^{1}\pi_* {\cal N}(-F -{\cal C})=0$
which follows from the explicit form of ${\cal N}(-F -{\cal C})$ in eq.~\eqref{3.3.14}.
Indeed, by definition,
the sheaf $R^{1}\pi_* {\cal N}(-F -{\cal C})$ is generated at each point $p$ 
on $\sigma$ by the cohomology of the fiber 
$H^1(F_p, {\cal N}(-F -{\cal C})|_{F_p})$, 
where $F_p$ is the elliptic fiber over $p$. From eq.~\eqref{3.3.14} and intersection 
numbers~\eqref{3.1.5}
it follows that the degree of ${\cal N}(-F -{\cal C})|_{F_p}$ is $3$ which is positive. 
Then it follows from the Kodaira vanishing theorem~\cite{GH} that 
$H^1(F_p, {\cal N}(-F -{\cal C})|_{F_p})=0$. Thus, $R^{1}\pi_* {\cal N}(-F -{\cal C})$
is the zero sheaf. To continue, 
from eqs.~\eqref{3.3.14} and~\eqref{3.1.15} we find that 
\begin{equation}
\pi_* {\cal N}(-F-{\cal C})={\cal O}(-6) \oplus {\cal O}(-8)\oplus{\cal O}(-9). 
\label{3.3.19}
\end{equation}
Since $h^1 ({\mathbb P}^1, {\cal O}(-i))=i-1$ for positive $i$, we find that the dimension 
of $W_1$ is 
\begin{equation}
{\rm dim} W_1 =5+7+8=20. 
\label{3.3.20}
\end{equation}
Moreover, the decomposition~\eqref{3.3.19} allows us to parametrize the elements of $W_1$ 
in terms of the differentials on $\sigma$.
Let $B_{-i} \in H^1(\sigma, {\cal O}(-i))$, $i= 6$, $8$, $9$ be the differentials 
on $\sigma$. Let $b_{-i} =\pi^* B_{-i}$ be their pullback to $d P_9$. 
$B_{-i}$ are elements of $H^1(d P_9, {\cal O}_{d P_9}(- i F))$. To construct 
an element $w_1 \in W_1$ we need to multiply $\pi^* B_{-6}$ by a section 
of ${\cal O}_{d P_9}(3 \sigma)$,  $\pi^* B_{-8}$ by a section of 
${\cal O}_{d P_9}(3 \sigma+2 F)$ and  $\pi^* B_{-9}$ by a section of 
${\cal O}_{d P_9}(3 \sigma+3 F)$. We can choose these sections to be $z$, $x$ and $y$. 
Thus, $w_1 \in W_1$ can be parametrized as 
\begin{equation}
w_1 =b_{-6} z +b_{-8}x +b_{-9}y. 
\label{3.3.21}
\end{equation}
Similarly, one can parametrize $w_2 \in W_2$. First, we note that 
\begin{eqnarray}
&&W_2=
H^1(d P_9, {\cal N}(-F)) \simeq H^1 (\sigma, \pi_* {\cal N}(-F)) \nonumber \\
&&={\cal O}(-1) \oplus {\cal O}(-3) \oplus {\cal O}(-4) 
\oplus {\cal O}(-5)  \oplus {\cal O}(-6)  \oplus {\cal O}(-7). 
\label{3.3.22}
\end{eqnarray}
The dimension of $W_2$ is then given by 
\begin{equation}
{\rm dim }W_2 =0+2+3+4+5+6=20. 
\label{3.3.23}
\end{equation}
Now an element $w_2 \in W_2$ can be written as follows
\begin{equation}
w_2 =c_3 z x + c_4 z y + c_5 x^2 + c_6 x y+ c_7 y^2, 
\label{3.3.24}
\end{equation}
where $c_{-j}=\pi^* C_{-j}$, $j=3$, $4$, $5$, $6$, $7$, where 
$C_{-j}$ are elements of $H^1(\sigma, {\cal O}(-j))$,
that is differentials on $\sigma$. The map $f_{{\cal C}}$ is a multiplication
of $w_1$ in eq.~\eqref{3.3.21} by ${\cal C}$ in eq.~\eqref{3.3.16}. 
The result of it must be an element $w_2$ in eq.~\eqref{3.3.24}. 
This multiplication can be organized in a $20 \times 20$ matrix depending 
on  $\{\psi_a, \phi_b, \chi_c\}$ in~\eqref{3.3.17}. 
We present some details of construction of this matrix in Appendix C. 
The determinant of this matrix is 
\begin{equation}
{\rm det} f_{{\cal C}}={\cal P}^4, 
\label{3.3.25}
\end{equation}
where ${\cal P}$ is a homogeneous polynomial of degree $5$
\begin{eqnarray}
&&{\cal{P}} =
\chi_1^2 \chi_3 \phi_3^2 -
\chi_1^2 \chi_2 \phi_3 \phi_4 -
2\chi_1 \chi_3^2  \phi_3 \phi_1 - \nonumber \\
&&\chi_1 \chi_2 \chi_3  \phi_3 \phi_2 +
\chi_2^2 \chi_3  \phi_1 \phi_3 +
\phi_4^2 \chi_1^3 -              \nonumber \\
&&2 \phi_2 \phi_4 \chi_3 \chi_1^2  +
\chi_1 \chi_3^2 \phi_2^2 +
3 \phi_1 \phi_4 \chi_1 \chi_2 \chi_3 + \nonumber \\
&&\phi_2 \chi_1 \phi_4 \chi_2^2 +
\phi_1^2 \chi_3^3 -
\phi_2 \chi_2 \phi_1 \chi_3^2-
\phi_4 \phi_1 \chi_2^3.
\label{3.3.26}
\end{eqnarray}
Note that ${\cal P}$ does not depend on $\psi_a$. 
Eqs.~\eqref{3.3.25}, \eqref{3.3.26} represent an
explicit equation in the moduli space of the vector bundle near 
which some number of (anti)-fundamental multiplets becomes light. 
Note that, this is not enough to generate massive SQCD in the free magnetic 
range since we need to know how many multiplets become light. 
We will analyze it later in this section. Before that, we will show that 
the reason why the (anti)-fundamental multiplets are massive away from the zero locus 
of ${\rm det} f_{{\cal C}}$ is precisely the Yukawa-type superpotential~\eqref{2.2.12}, \eqref{2.2.13}. 


\subsection{The Superpotential}


In this subsection, we will show that the exact sequence~\eqref{3.3.10}
which we can write as
\begin{equation}
0 \rightarrow H^0(\sigma, K_{\sigma}^{1/2} \otimes V|_{\sigma})
\rightarrow W_1\stackrel{f_{{\cal C}}}{\rightarrow} W_2
\rightarrow H^1(\sigma, K_{\sigma}^{1/2} \otimes V|_{\sigma})\rightarrow \dots
\label{3.4.1}
\end{equation}
can be interpreted as an algebraic geometry version of the superpotential~\eqref{2.2.12}, \eqref{2.2.13}. 
First, we will use the Serre duality to write 
\begin{equation}
H^1(\sigma, K_{\sigma}^{1/2} \otimes V|_{\sigma}) \simeq 
H^0(\sigma, K_{\sigma}^{1/2} \otimes V^{\vee}|_{\sigma})^{\vee}. 
\label{3.4.2}
\end{equation}
Second, from~\eqref{3.4.1} it follows that 
$H^0(\sigma, K_{\sigma}^{1/2} \otimes V|_{\sigma})$
is a subgroup of $W_1$. Similarly, $H^0(\sigma, K_{\sigma}^{1/2} \otimes V^{\vee}|_{\sigma})$ 
is a subgroup of $W_2^{\vee}$. When we multiply an element $w_1 \in W_1$ by $f_{{\cal C}}$  
we obtain an element $w_2 \in W_2$. This element can be paired up 
with an element of $W_2^{\vee}$ to produce a complex number. The map 
$f_{{\cal C}}$ depends on the vector bundle moduli and, hence, can be viewed as an element
in $H^1(\sigma, (V \otimes V^{\vee})|_{\sigma})$. 
Thus, the sequence~\eqref{3.4.1} gives a natural map 
\begin{equation}
H^0(\sigma, K_{\sigma}^{1/2} \otimes V|_{\sigma}) 
\otimes 
H^1(\sigma, (V \otimes V^{\vee})|_{\sigma})
\otimes
H^0(\sigma, K_{\sigma}^{1/2} \otimes V^{\vee}|_{\sigma}) 
\to {\mathbb C}. 
\label{3.4.3}
\end{equation}
This map is exactly the superpotential as explained at the end 
of subsection 3.4. 


\subsection{Examples of SQCD}


In this final subsection, we will give examples of SQCD in the free magnetic range
within the framework of the {\bf Example} given in subsection 4.2.
For this we need to understand how many (anti)-fundamental
flavors become
light near the locus ${\rm det} f_{{\cal C}}=0$ in eqs.\eqref{3.3.25}, \eqref{3.3.26}.
This number is the dimension of the kernel of the matrix $f_{{\cal C}}$. 
At any point in the moduli space where ${\rm det} f_{{\cal C}}=0$
the rank of the matrix $f_{{\cal C}}$ is less than $20$. Note that the rank 
changes as we move in the zero locus of  ${\rm det} f_{{\cal C}}$. 
Let $r$ be the rank of $f_{{\cal C}}$ at some point in the moduli space. 
Then the dimension of the kernel of $f_{{\cal C}}$ is simply $20-r$. 
Unfortunately, a detailed study of the rank of $f_{{\cal C}}$ in different regimes
in the moduli space requires a hard numeric work. 
However, for some values of the moduli $\psi_a, \phi_b, \chi_c$
the matrix $f_{{\cal C}}$ simplifies and one can prove the existence 
of subspaces where a certain specific number of flavors becomes light. 
It will be enough to present examples of SQCD in the free magnetic range. 

Let us study the subspace of ${\mathbb P}^{12}$ where 
$\psi_a =0$, $a=1, \dots, 6$.
Then one can show that it is possible to arrange the rows and columns
in such a way that
the matrix $f_{{\cal C}}$ becomes block-diagonal
with four identical $5\times 5$ blocks of the form 
\begin{equation}
M ={\ }{\ }
\bordermatrix{    
& {\ }     & {\ }    & {\ }    & {\ }   & {\ }   \cr
& \phi_1   & \phi_2   & \phi_3 & \phi_4 & 0      \cr
& 0        & \phi_1   & \phi_2 & \phi_3 & \phi_4 \cr
& \chi_1   & \chi_2   & \chi_3 & 0      & 0      \cr
& 0        & \chi_1   & \chi_2 & \chi_3 & 0      \cr
& 0        & 0        & \chi_1 & \chi_2 & \chi_3 \cr}.
\label{m1}
\end{equation}
Note that the determinant of $M$ is precisely the polynomial ${\cal P}$ 
in eq.~\eqref{3.3.26}. In other words, the determinant of the whole 
matrix $f_{{\cal C}}$ is the determinant of $M$ raised to the power four. 
It is easy to see that setting, for example, $\phi_1=\chi_1=\chi_2=0$ reduces the rank of the 
matrix $M$ by one. Since $f_{{\cal C}}$ consists of four blocks of $M$ 
the rank of $f_{{\cal C}}$ at this locus drops by four. This proves that there exist a 
subvariety ${\mathbb L}_1 \subset{\mathbb P}^{12}$ containing the subspace 
\begin{equation}
\psi_a=\phi_1=\chi_1=\chi_2=0, \quad a=1, \dots, 6,
\label{m2}
\end{equation}
where the rank of the matrix $f_{{\cal C}}$ drops by four. 
Similarly, it is not difficult to prove that there exists a subvarity 
${\mathbb L}_2 \in {\mathbb P}^{12}$ where the rank of $M$ drops by two 
and the rank of $f_{{\cal C}}$ drops by eight. For example, the following 
subspace is contained in ${\mathbb L}_2$
\begin{equation}
\psi_a=\phi_1=\phi_2=\chi_1=\chi_2=0, \quad a=1, \dots, 6.
\label{m3}
\end{equation}
Of course, the subvarieties ${\mathbb L}_1$ and ${\mathbb L}_2$
are much wider than the their subspaces specified in eqs.~\eqref{m2} and~\eqref{m3}.
However, for our purposes it is enough to establish that   
${\mathbb L}_1$ and ${\mathbb L}_2$ are non-empty. 
It is very likely that by turning on the moduli $\psi_a$ one can achieve that 
the rank of $f_{{\cal C}}$ drops by any number between $4$ and $8$.
Now we give some examples 
of SQCD in the free magnetic range. 
\begin{itemize}
\item 
Let us choose the low-energy gauge group to be
\begin{equation}
\Gamma_{{\cal S}^{\prime}}=SU(3).
\label{3.5.1.2}
\end{equation}
The free magnetic range for the gauge group $SU(3)$ is given by 
\begin{equation}
4 \leq N_f < \frac{9}{2}. 
\label{3.5.2}
\end{equation}
We see that $N_f=4$ is a solution to~\eqref{3.5.2}. 
We showed above that there exists a subvariety ${\mathbb L}_1$ in the moduli space
where the rank of $f_{{\cal C}}$ drops by four. This means that dimension 
of the kernel of $f_{{\cal C}}$ is four. This, in turn, means that near 
${\mathbb L}_1$ we have exactly four fundamental flavors.
Thus, 
near a generic point of the subvariety ${\mathbb L}_1$ 
we 
generate SQCD in the free magnetic range with 
\begin{equation}
N_c=3,\quad N_f=4. 
\label{3.5.2.1}
\end{equation}
\item 
Let us now choose the low-energy gauge group to be
\begin{equation}
\Gamma_{{\cal S}^{\prime}}=SU(6).
\label{3.5.2.2}
\end{equation}
The free magnetic range for the gauge group $SU(6)$ is given by 
\begin{equation}
7 \leq N_f < 9. 
\label{3.5.3}
\end{equation}
From our discussion earlier in this subsection
we know that there exists a subvariety ${\mathbb L}_2$ in the moduli space
where the rank of the matrix $f_{{\cal C}}$ drops by eight. Hence, the dimension 
of the kernel of $f_{{\cal C}}$ becomes eight. Thus, near ${\mathbb L}_2$
we have exactly eight fundamental flavors. 
This ways we generate SQCD in the free magnetic range with 
\begin{equation}
N_c=6,\quad N_f=8. 
\label{3.5.3.1}
\end{equation}
\end{itemize}

Note that the fact that ${\rm det} f_{{\cal C}}$ is given by a polynomial
of high degree is rather helpful in generating a suitable number of flavors. 

Clearly, using the technics presented in this paper, 
one can construct many other examples of SQCD on F-theory seven-branes
and find the regimes in the moduli
space where the number of flavors
is in the free magnetic range.


\section{Conclusion}


In this paper, we address the question of realizing dynamically SUSY 
breaking SQCD~\cite{ISS} in F-theory. Our starting point is the field theory 
on the intersecting seven-branes obtained by Beasley, Heckman and Vafa in~\cite{BHV}. 
In our model, one of the seven-branes realizes ${\cal N}=1$, 
$SU(N_c)$ supersymmetric Yang-Mills theory. 
The other one contributes vector bundle moduli.
Finally, the matter fields in the (anti)-fundamental representation of $SU(N_c)$ 
comes the intersection. These matter fields have a quadratic superpotential 
with the mass matrix depending on the vector bundle moduli. 
In order to obtain 
SUSY breaking SQCD in the free magnetic range one has to move to a certain regime
in the moduli space where an appropriate number of the matter fields becomes light. 
Conceptually, this is similar to analyzing how many 
Higgs multiplets one has in heterotic standard models 
of~\cite{Braun1, Braun2, Braun3, Braun4, Ron1, Ron2}.
For example, in the model of~\cite{Ron1} one can have zero, one or two Higgs multiplets 
depending on the location in the moduli space. 
Though in this paper, for concreteness, we work in the context of some 
specific choices of the type of the $ADE$-singularity 
and of the geometric data, our method has, of course, a wider
applicability.

A natural question which arises is whether it is possible to
generate 
the mass term for the (anti)-fundamental multiplets not by vector bundle moduli 
but by $D1$- or $D3$-brane instantons.
The mass obtained this way will be exponentially 
suppressed by the volume of the Euclidean $D$-brane. This idea of generating a small 
mass was used recently in other contexts in~\cite{Uranga, Cvetic, Shamit, Franco, Richter}
(see also~\cite{Lust, Lerda, Lerda1, Lerda2} for similar calculations).
If this Euclidean $D1$-brane (or $D3$-brane)
intersects the space filling branes, which are the seven-branes
in our case, in general, there are fermionic zero modes due to Ganor's strings~\cite{Ganor} 
stretched between the $D1$- (or $D3$-) and the space-filling branes. These instanton zero 
modes will couple to the (anti)-fundamental matter fields $Q$ and $\tilde{Q}$. 
Hence, upon integration of these Ganor's zero modes one can generate 
a non-perturbative superpotential for $Q$ and $\tilde{Q}$.
One can approach this problem by first generating 
a massless SQCD and then showing that one can 
produce the mass term by $D1$-
or $D3$-brane instantons intersecting the seven-branes. 
It would be interesting to explore this in the future.


\section{Acknowledgements}

The author is very grateful to Chris Beasley for 
explanations of the results 
of~\cite{BHV} and for interesting discussions. 
The author is also very grateful to
Mike Schulz and Tony Pantev for discussions.
Research at Perimeter Institute for Theoretical Physics is supported in
part by the Government of Canada through NSERC and by the Province of
Ontario through MRI.

\appendix
\section{The Twist on the Surface}

In this appendix we will review the twisting procedure to obtain a theory 
on ${\mathbb R}^{3, 1}\times {\cal S}$, where ${\cal S}$ is a compact Kahler 
surface over which we wrap the seven-branes. 

We start with the maximally supersymmetric theory 
on ${\mathbb R}^{3, 1}\times {\mathbb C}^2$.
The symmetry of this theory is $SO(7, 1) \times U(1)_{R}$.
In addition to the eight-dimensional gauge field, this theory 
contains a complex scalar $\phi, \bar \phi$ and two fermions $\Psi_{\pm}$ 
transforming under $SO(7, 1) \times U(1)_{R}$ as 
\begin{equation}
\left({\bf S}_+, \frac{1}{2}\right)
\label{B1} 
\end{equation}
and 
\begin{equation}
\left({\bf S}_-, -\frac{1}{2}\right)
\label{B2} 
\end{equation}
respectively. Here by ${\bf S}_{\pm}$ we denoted the positive and negative chirality 
representations of $SO(7, 1)$. This theory is invariant under two supersymmetries 
whose parameters $\epsilon_{\pm}$ 
transform in the same way as $\Psi_{\pm}$.~\footnote{The simplest way to see these results is 
to recall that this theory can be obtained by compactifying the ten-dimensional 
supersymmetric Yang-Mills theory to eight dimensions.}
Our aim is to obtain a theory on ${\mathbb R}^{3, 1}\times {\cal S}$
whose symmetry is reduced to 
$SO(3, 1) \times SO(4) \times U(1)_{R}$. 
Here $SO(3, 1)$ is the Lorentz group in four dimensions and 
$SO(4)$ is the structure group of the tangent bundle of ${\cal S}$. 
The parameters $\epsilon_{\pm}$ decompose as follows
\begin{equation}
\epsilon_+ \in \left({\bf S}_+, \frac{1}{2}\right) \to 
\left[({\bf 2}, {\bf 1}), ({\bf 2}, {\bf 1}), \frac{1}{2}\right]\oplus
\left[({\bf 1}, {\bf 2}), ({\bf 1}, {\bf 2}), \frac{1}{2}\right] 
\label{B3}
\end{equation}
and 
\begin{equation}
\epsilon_- \in \left({\bf S}_-, -\frac{1}{2}\right) \to 
\left[({\bf 2}, {\bf 1}), ({\bf 1}, {\bf 2}), -\frac{1}{2}\right]\oplus
\left[({\bf 1}, {\bf 2}), ({\bf 2}, {\bf 1}), -\frac{1}{2}\right], 
\label{B4}
\end{equation}
where by $({\bf 2}, {\bf 1})$ we denote the left-handed spinor 
of $SO(3, 1)$ (or  $SO(4)$ depending on its position in the square brackets)
and by $({\bf 1}, {\bf 2})$ we denote the right-handed spinor. 

The twisting procedure is described by an embedding of $U(1)_R$ 
into $SO(4)$. In fact, since ${\cal S}$ is Kahler, its structure group is reduced 
to $U(2)$. Thus, we need to specify how $U(1)_R$ is embedded in $U(2)$. 
It was argued in~\cite{BHV} that the unique choice up to isomorphism is the 
twist under which $U(1)_R$ is embedded into the center of $U(2)$. Let $J$ 
be the generator of this central $U(1)$. We can normalize $J$ in such a way that
under the reduction of $SO(4)$ to $U(2)$ the spinors of $SO(4)$ 
transform as 
\begin{equation}
({\bf 2}, {\bf 1})\to {\bf 2}_0, \qquad ({\bf 1}, {\bf 2})\to {\bf 1}_{1}\oplus {\bf 1}_{-1}, 
\label{B5}
\end{equation}
where the subscripts denote the charge under $J$. Then from eqs.~\eqref{B3}, \eqref{B4} 
and~\eqref{B5} it follows that to preserve four supercharges in four dimensions 
the new $U(1)$ generator has to be chosen to be 
\begin{equation}
J_{top} =J\pm 2 R. 
\label{B6}
\end{equation}
It is easy to see that either choice of the sign leads 
to an isomorphic twist. We will choose $J_{top} =J + 2 R$. 
Let us check that we indeed obtain four supercharges. Under $SO(3, 1) \times U(2)$ 
the supersymmetry generators transform as 
\begin{eqnarray}
&&\left[({\bf 2}, {\bf 1}), {\bf 2}_1\right]\oplus 
\left[({\bf 1}, {\bf 2}) \otimes ({\bf 1}_2 \oplus {\bf 1}_0)\right], \nonumber\\
&&\left[({\bf 1}, {\bf 2}), {\bf 2}_{-1}\right]\oplus 
\left[({\bf 2}, {\bf 1}) \otimes ({\bf 1}_0 \oplus {\bf 1}_{-2})\right].
\label{B7}
\end{eqnarray}
Four-dimensional supercharges have to be scalars on ${\cal S}$ and, hence, 
correspond to the terms 
$({\bf 1}, {\bf 2})\otimes  {\bf 1}_0$ and 
$({\bf 2}, {\bf 1})\otimes  {\bf 1}_0$. 

Now let us find how the scalars $\phi$ and $\bar \phi$ transform is the twisted theory. 
Before the twist they transformed as ${\bf 1}\otimes {\bf 1}_{\pm 1}$ under 
$SO(3, 1) \times U(2)$. According to~\eqref{B6}, after the twist they transform as 
${\bf 1}\otimes {\bf 1}_{\pm 2}$. Let us interpret it geometrically. 
We fix conventions that the central $U(1)$ of $U(2)$ acts on vectors 
of the holomorphic vector bundle with charge $+1$. Then it acts on holomorphic 
differential forms with charge $-1$. Let $s^{m}, {\bar s}^{\bar m}$ be holomorhic 
and antiholomorphic coordinates on ${\cal S}$. Then $d s^m$ has charge $-1$ and 
$d {\bar s}^{\bar m}$ has charge $+1$. Therefore, $\phi$ and $\bar \phi$
become the following differential forms 
\begin{equation}
\phi=\phi_{m n} d s^m d s^n, \qquad 
\bar \phi=\bar \phi_{{\bar m} {\bar n}}  {\bar s}^{\bar m} {\bar s}^{\bar n}.
\label{B8}
\end{equation}
Similarly, one can analyze the fermions. The results are summarized in subsection 2.2. 

\section{The Twist on the Curve}


To discuss the theory on the intersection curve $\Sigma$ we start with the untwisted theory
on ${\mathbb R}^{1, 1}$. This theory preserves eight supercharges and has a pair 
of complex scalars $(\sigma, \bar \sigma^c)$ forming a doublet of the R-symmetry group 
$SU(2)_R$ and a chiral fermion (we choose its chirality to be negative) which 
transforms as ${\bf 4}^{\prime}$ of the Lorentz group $SO(5, 1)$. 
The supersymmetry generators transform as ${\bf 4}^{\prime}\otimes {\bf 2}$ 
of $SO(5, 1)\times SU(2)_R$. In order to twist we reduce $SO(5, 1)$ to 
$SO(3, 1)\times U(1)$ where $SO(3, 1)$ is the Lorentz group in four dimensions
and $U(1)$ is the structure group of the tangent bundle of $\Sigma$. The representations
${\bf 4}^{\prime}$ of $SO(5, 1)$ decomposes under $SO(3, 1)\times U(1)$ as 
\begin{equation}
{\bf 4}^{\prime} \to 
\left[{\bf (2, 1)},-\frac{1}{2}\right]\oplus \left[{\bf (1, 2)},\frac{1}{2}\right]. 
\label{C1}
\end{equation}
In addition, ${\bf 2}$ of $SU(2)_R$ decomposes to the Cartan subgroup
$U(1)_R$ as
as 
\begin{equation}
{\bf 2}\to {\bf 1}_1 \oplus {\bf 1}_{-1}. 
\label{C2}
\end{equation}
The twisting procedure is a specification of a homomorphism from $U(1)_R$  
to the structure group $U(1)$. Let $J$ be the generator of the structure group 
$U(1)$. To preserve ${\cal N}=1$ supersymmetry four supercharges must become 
scalars on $\Sigma$. This requires that the generator of the twisted $U(1)$ be 
\begin{equation}
J_{top}=J \pm \frac{1}{2}R. 
\label{C3}
\end{equation}
Either choice of the sign leads to an isomorphic twist. We will choose the sign to be minus.

Let us see what happens to the scalars $(\sigma, \bar \sigma^c)$
under this twist. Since they are scalars under $SO(5, 1)$ they have $J=0$. 
On the other hand they carry the charge $\pm 1$ under $U(1)_R$. Thus, after 
the twist their charges become $\mp \frac{1}{2}$.
This means that they become spinors on $\Sigma$. More precisely, 
\begin{equation}
\sigma \in K_{\Sigma}^{1/2}, \qquad {\bar \sigma}^c \in {\bar K}_{\Sigma}^{1/2}.
\label{C4}
\end{equation}
Since the fermions do not transform under $SU(2)_R$, $J_{top}=J$ and 
the twist does not affect their geometric properties. 
The full spectrum is summarized in subsection 2.2. Of course, the above fields 
are charged under the gauge group. However, it is not affected by the twist and we have omitted 
the gauge group in this discussion. 


\section{Construction of the Matrix $f_{{\cal C}}$}


In this appendix we will present some details of construction of the $20 \times 20$ 
matrix $f_{{\cal C}}$ in subsection 4.2. 

The matrix $f_{{\cal C}}$ provides a linear map between two
$20$-dimensional spaces $W_1$ and $W_2$ given by
\begin{eqnarray}
W_1 & = &H^1(d P_9, {\cal O}_{d P_9}(3 \sigma-6 F))
\simeq H^1(\sigma, {\cal O}(-6)\oplus {\cal O}(-8) \oplus {\cal O}(-9)), \nonumber \\
W_2 & = &H^1(d P_9, {\cal O}_{d P_9}(6 \sigma- F)) \nonumber \\
 & \simeq &
 H^1(\sigma, {\cal O}(-1)\oplus {\cal O}(-3) \oplus {\cal O}(-4)
\oplus
{\cal O}(-5)\oplus {\cal O}(-6) \oplus {\cal O}(-7)). 
\label{A1}
\end{eqnarray}
The elements $w_1 \in W_1$ and $w_2 \in W_2$ have been parametrized as follows 
\begin{eqnarray}
&& w_1 =b_{-6} z + b_{-8} x + b_{-9} y, \nonumber\\
&& w_2 =c_{-3} z x + c_{-4} z y + c_{-5} x^2  + c_{-6} x y + c_{-7} y^2. 
\label{A2}
\end{eqnarray}
In these expressions, $b_{-i}$ and $c_{-j}$ are the pullback to $d P_9$ 
of the differentials on $\sigma \simeq {\mathbb P}^1$
\begin{equation}
b_{-i} =\pi^* B_{-i}, \quad c_{-j}= \pi^* C_{-j}, 
\label{A3}
\end{equation}
where $B_{-i}\in H^1({\sigma}, {\cal O}(-i))$, $i=6$, $8$, $9$ and 
$C_{-j}\in H^1({\sigma}, {\cal O}(-j))$, $j=3$, $4$, $5$, $6$, $7$. 
Furthermore, $z$, $x$, $y$ are the variables in the Weierstrass equation 
\begin{equation}
y^2 z =x^3 + f x z^2 +g z^3. 
\label{A4}
\end{equation}
They are sections of the following line bundles on $d P_9$
\begin{equation}
z\in H^0(d P_9, {\cal O}_{d P_9} (3 \sigma)), \quad 
x\in H^0(d P_9, {\cal O}_{d P_9} (3 \sigma+2 F)), \quad 
y\in H^0(d P_9, {\cal O}_{d P_9} (3 \sigma+3 F)).
\label{A5}
\end{equation}
Note that each term in the sum in $w_1$ and $w_2$ in eqs.~\eqref{A2}
is an element of $H^1(d P_9, {\cal O}_{d P_9}(3 \sigma-6 F))$ and 
$H^1(d P_9, {\cal O}_{d P_9}(6\sigma- F))$ respectively. The map between 
$w_1$ and $w_2$ is given by multiplication by an element of 
$H^0 (d P_9, {\cal O}_{d P_9}(3\sigma +5 F))$ which we write as
\begin{equation}
{\cal C}=a_5 z+a_3 x +a_2 y, 
\label{A6}
\end{equation}
where $a_k \in H^0(d P_9, \pi^* {\cal O}(k))$, $k=1$, $3$, $5$. This means 
that $a_k=\pi^* A_k$, where $A_k$ is a homogeneous polynomial of degree $k$ on $\sigma$. 
In subsection 4.2 we introduce homogeneous coordinates $(u, v)$ on $\sigma$ and 
parametrized $A_k$ as follows
\begin{eqnarray}
&&A_5 = \psi_1 u^5 + \psi_2 u^4 v + \psi_3 u^3 v^2+ \psi_4 u^2 v^3+\psi_5   u v^4 +\psi_6 v^5, 
\nonumber \\
&&A_3 = \phi_1 u^3 + \phi_2 u^2 v + \phi_3 u v^2+ \phi_4 v^3, 
\nonumber \\
&&A_2 = \chi_1 u^2 + \chi_2 u v + \chi_3 v^3, 
\label{A7}
\end{eqnarray}
where $\{\psi_a, \phi_b, \chi_c\}$ are the {\it projective} vector bundle moduli. 
To simplify our notation, we will remove the pullback symbol $\pi^*$ and identify 
$b_{-i}=B_{-i}$, $c_{-j}=C_{-j}$ and $a_{-k}=A_{-k}$
and view the coefficients $b_{-i}$, $c_{-j}$ and $a_{k}$ in eqs.~\eqref{A2} and~\eqref{A6}
as differentials and polynomials on $\sigma$. 

Suppressing for the time being the coefficients $b_{-i}$ and $c_{-j}$ we see that 
$W_1$ us spanned by the the following basis blocks 
\begin{equation}
z, x, y. 
\label{A8}
\end{equation}
Similarly, $W_2$ is spanned by the basis blocks 
\begin{equation}
z x, x y, x^2, x y, y^2. 
\label{A9}
\end{equation}
Now we multiply $w_1$ in eq.~\eqref{A2} by ${\cal C}$ in eq.~\eqref{A6} and expand
the answer in basis elements in~\eqref{A9}. We obtain the following matrix $f_{{\cal C}}$
\begin{equation}
f_{{\cal C}}={\ }{\ }
\bordermatrix{    & z   & x   & y  \cr
              xz  & a_3 & a_5 & 0  \cr
              yz  & a_2 & 0   & a_5 \cr
              x^2 & 0   & a_3 & 0   \cr
              xy  & 0   & a_2 & a_3 \cr
              y^2 & 0   & 0   & a_2 \cr}.
\label{A10}
\end{equation}
The matrix $f_{{\cal C}}$ is written in a block form where each block is a
$(j-1) \times (i-1)$ matrix for $j=3$, $4$, $5$, $6$, $7$ and 
$i=6$, $8$, $9$. Now we can compute each block by expanding $a_k$ 
in the coordinates $(u, v)$ as in~\eqref{A7}. For example, let us compute the 
$z - z x$ block $a_3$. That is, we want to compute the map
\begin{equation}
H^1(d P_9, {\cal O}_{d P_9}(3 \sigma -6 F))|_{b_{-6}}
\stackrel{a_3}{\rightarrow} H^1(d P_9, {\cal O}_{d P_9}(6 \sigma -F))|_{c_{-3}}. 
\label{A11}
\end{equation}
The map is a multiplication by $a_3$ in~\eqref{A7}. Note that
\begin{equation}
h^1 (d P_9, {\cal O}_{d P_9}(3 \sigma -6 F))|_{b_{-6}}=h^1(\sigma, {\cal O}(-6))=5
\label{A12}
\end{equation}
and 
\begin{equation}
h^1(d P_9, {\cal O}_{d P_9}(6 \sigma -F))|_{c_{-3}}=h^1(\sigma, {\cal O}(-3))=2. 
\label{A13}
\end{equation}
Therefore, the block $z - z x$ in the matrix~\eqref{A10} is a $2 \times 5$ matrix. 
To construct $a_3$ in~\eqref{A11} and~\eqref{A10} we use the Serre duality to identify
\begin{equation}
H^1(\sigma, {\cal O}(-6)=H^0(\sigma, {\cal O}(4))^{\vee}
\label{A14}
\end{equation}
and 
\begin{equation}
H^1(\sigma, {\cal O}(-3)=H^0(\sigma, {\cal O}(1))^{\vee}
\label{A15}
\end{equation}
Let us introduce the two-dimensional linear space
\begin{equation}
\hat{V}=H^0(\sigma, {\cal O}(1)). 
\label{A16}
\end{equation}
It is parametrized by the linear functions on $\sigma$. That is, by the 
projective coordinates $(u, v)$. Similarly, we introduce the dual vector space 
\begin{equation}
\hat{V}^{\vee}=H^0(\sigma, {\cal O}(1))^{\vee}
\label{A17}
\end{equation}
and parametrize it by the dual basis $(u^*, v^*)$, where
\begin{equation}
u^* u=1, \quad v^* v=1, \quad u^* v= u v^*=0. 
\label{A18}
\end{equation}
Then from eq.~\eqref{A14} it follows that $H^1(\sigma, {\cal O}(-6))$ is 
spanned by the following basis
\begin{equation}
\{ u^{*4}, u^{* 3} v^*, u^{* 2} v^{*2},  u^{* } v^{*3},  v^{*4}\}. 
\label{A19}
\end{equation}
Similarly, $H^1(\sigma, {\cal O}(-3))\simeq \hat{V}^{\vee}$ is spanned by 
\begin{equation}
\{u^*, v^*\}. 
\label{A20}
\end{equation}
The coefficient $a_3$ is a map between~\eqref{A19} and~\eqref{A20}. 
Multiplying basis elements in~\eqref{A19} by $a_3$ in~\eqref{A7} and
using relations~\eqref{A18} we obtain the following $2 \times 5$ matrix
\begin{equation}
\bordermatrix{    & u^{*4}  & u^{*3}v^{*} & u^{*2}v^{*2}  & u^{*}v^{*3} &
v^{*4} \cr
              u^* & \phi_1  & \phi_2      & \phi_3        & \phi_4      &
0      \cr
              v^* & 0       & \phi_1      & \phi_2        & \phi_3      &
\phi_4  \cr}.
\label{A21}
\end{equation}
Continuing this way, one can build up the complete 
matrix $f_{{\cal C}}$. The determinant of this matrix is given 
in eq.~\eqref{3.3.26}. 



\begin{thebibliography}{99}


\bibitem{Vafa}
C. Vafa, ``Evidence for F-Theory,''
Nucl.Phys. B469 (1996) 403-418 [arXiv:hep-th/9602022].

\bibitem{MVafa1}
D. R. Morrison and C. Vafa,
``Compactifications of F-Theory on Calabi--Yau Threefolds -- I,''
Nucl.Phys. B473 (1996) 74-92 [arXiv:hep-th/9602114].

\bibitem{Mvafa2}
D. R. Morrison and C. Vafa,
``Compactifications of F-Theory on Calabi--Yau Threefolds -- II,''
Nucl.Phys. B476 (1996) 437-469 [arXiv:hep-th/9603161].

\bibitem{Vafasix}
M. Bershadsky, K. Intriligator, S. Kachru, D.R. Morrison, V. Sadov and C. Vafa,
``Geometric Singularities and Enhanced Gauge Symmetries,''
Nucl.Phys. B481 (1996) 215-252 [arXiv:hep-th/9605200].

\bibitem{FMW}
R. Friedman, J. Morgan and E. Witten, 
``Vector Bundles And F Theory,''
Commun.Math.Phys. 187 (1997) 679-743 [arXiv:hep-th/9701162].

\bibitem{Tony}
M. Bershadsky, A. Johansen, Tony Pantev and V. Sadov,
``On Four-Dimensional Compactifications of F-Theory,''
Nucl.Phys. B505 (1997) 165-201 [arXiv:hep-th/9701165].

\bibitem{Curio}
G. Curio and R. Donagi, 
``Moduli in N=1 heterotic/F-theory duality,''
Nucl.Phys. B518 (1998) 603-631 [arXiv:hep-th/9801057].

\bibitem{Andreas1}
B. Andreas and G. Curio, 
``On discrete Twist and Four-Flux in N=1 heterotic/F-theory compactifications,''
Adv.Theor.Math.Phys. 3 (1999) 1325-1413 [arXiv:hep-th/9908193].

\bibitem{Andreas2}
B. Andreas and G. Curio, 
``Horizontal and Vertical Five-Branes in Heterotic/F-Theory Duality,''
JHEP 0001 (2000) 013 [arXiv:hep-th/9912025].

\bibitem{Diaconescu0}
D.-E. Diaconescu and G. Rajesh, 
``Geometrical Aspects of Fivebranes in Heterotic/F-Theory Duality in Four Dimensions,''
JHEP 9906 (1999) 002 [arXiv:hep-th/9903104].

\bibitem{Braun1}
V. Braun, Y.-H. He, B. A. Ovrut and T. Pantev, 
``A Heterotic Standard Model,''
Phys.Lett. B618 (2005) 252-258 [arXiv:hep-th/0501070].

\bibitem{Braun2}
V. Braun, Y.-H. He, B. A. Ovrut and T. Pantev,
``A Standard Model from the E8 x E8 Heterotic Superstring,''
JHEP 0506 (2005) 039 [arXiv:hep-th/0502155].

\bibitem{Braun3}
V. Braun, Y.-H. He, B. A. Ovrut and T. Pantev,
``Vector Bundle Extensions, Sheaf Cohomology, and the Heterotic Standard Model,''
Adv.Theor.Math.Phys. 10 (2006) 4 [arXiv:hep-th/0505041].

\bibitem{Braun4}
V. Braun, Y.-H. He, B. A. Ovrut and T. Pantev,
``The Exact MSSM Spectrum from String Theory,''
JHEP 0605 (2006) 043 [arXiv:hep-th/0512177]. 

\bibitem{Ron1}
V. Bouchard and R. Donagi, 
``An SU(5) Heterotic Standard Model,''
Phys.Lett. B633 (2006) 783-791 [arXiv:hep-th/0512149].

\bibitem{Ron2}
V. Bouchard and R. Donagi, 
``On heterotic model constraints,''
arXiv:0804.2096.

\bibitem{DougK}
M. R. Douglas and S. Kachru, 
``Flux Compactification,''
Rev.Mod.Phys.79:733-796,2007 [arXiv:hep-th/0610102].

\bibitem{Denef1}
F. Denef, 
``Les Houches Lectures on Constructing String Vacua,''
arXiv:0803.1194. 

\bibitem{Baumann}
D. Baumann, A. Dymarsky, I. R. Klebanov and L. McAllister, 
``Towards an Explicit Model of D-brane Inflation,''
JCAP 0801:024,2008 [arXiv:0706.0360].

\bibitem{Lust0}
M. Haack, R. Kallosh, A. Krause, A. Linde, D. Lust and M. Zagermann, 
``Update of $D3/D7$-Brane Inflation on $K3 \times T^2/Z_2$,''
arXiv:0804.3961.

\bibitem{Martijn}
R. Donagi and M. Wijnholt, 
``Model Building with F-Theory,''
arXiv:0802.2969.

\bibitem{BHV}
C. Beasley, J. J. Heckman and C. Vafa,
``GUTs and Exceptional Branes in F-theory - I,''
arXiv:0802.3391.

\bibitem{Tatar}
H. Hayashi, R. Tatar, Y. Toda, T. Watari and M. Yamazaki, 
``New Aspects of Heterotic--F Theory Duality,''
arXiv:0805.1057.

\bibitem{BHV2}
C. Beasley, J. J. Heckman and C. Vafa, {\it work in progress}.

\bibitem{ISS}
K. Intriligator, N. Seiberg and D. Shih,
``Dynamical SUSY Breaking in Meta-Stable Vacua,''
JHEP 0604 (2006) 021 [arXiv:hep-th/0602239].

\bibitem{Diaconescu}
D.-E. Diaconescu, B. Florea, S. Kachru and P. Svrcek, 
``Gauge-Mediated Supersymmetry Breaking in String Compactifications,''
JHEP 0602 (2006) 020 [arXiv:hep-th/0512170]. 

\bibitem{BBO1}
V. Braun, E. I. Buchbinder and B. A.Ovrut,
``Dynamical SUSY Breaking in Heterotic M-Theory,''
Phys.Lett. B639 (2006) 566-570 [arXiv:hep-th/0606166].

\bibitem{BBO2}
V. Braun, E. I. Buchbinder and B. A.Ovrut,
``Towards Realizing Dynamical SUSY Breaking in Heterotic Model Building,''
JHEP 0610 (2006) 041 [arXiv:hep-th/0606241].

\bibitem{Greene}
J. Distler and B. R. Greene, 
``Some Exact Results on the Superpotential from Calabi-Yau Compactifications,''
Nucl.Phys.B309:295,1988. 

\bibitem{Witten99}
E. Witten, 
``World-Sheet Corrections Via D-Instantons,''
JHEP 0002 (2000) 030 [arXiv:hep-th/9907041].

\bibitem{BDO2}
E. I. Buchbinder, R. Donagi and B. A. Ovrut, 
``Superpotentials for Vector Bundle Moduli,''
Nucl.Phys. B653 (2003) 400-420 [arXiv:hep-th/0205190].

\bibitem{BDO3}
E. I. Buchbinder, R. Donagi and B. A. Ovrut, 
``Vector Bundle Moduli Superpotentials in Heterotic Superstrings and M-Theory,''
JHEP 0207 (2002) 066 [arXiv:hep-th/0206203].

\bibitem{Donagi}
R. Donagi, ``Principal bundles on elliptic fibrations,''
Asian J. Math. Vol. 1 (June 1997), 214-223 [arXiv:alg-geom/9702002].

\bibitem{Seiberg1}
N. Seiberg,
``Exact Results on the Space of Vacua of Four Dimensional SUSY Gauge Theories,''
Phys.Rev. D49 (1994) 6857-6863 [arXiv:hep-th/9402044].

\bibitem{Seiberg2}
N. Seiberg,
``Electric-Magnetic Duality in Supersymmetric Non-Abelian Gauge Theories,''
Nucl.Phys. B435 (1995) 129-146 [arXiv:hep-th/9411149].

\bibitem{OO1}
H. Ooguri and Y. Ookouchi, 
``Meta-Stable Supersymmetry Breaking Vacua on Intersecting Branes,''
Phys.Lett. B641 (2006) 323-328 [arXiv:hep-th/0607183].

\bibitem{Franco1}
S. Franco, I. Garcia-Etxebarria and A. M. Uranga, 
``Non-supersymmetric Meta-stable Vacua from Brane Configurations,''
JHEP 0701 (2007) 085 [arXiv:hep-th/0607218].

\bibitem{Bena}
I. Bena, E. Gorbatov, S.Hellerman, N. Seiberg and D. Shih, 
``A Note on (Meta)stable Brane Configurations in MQCD,''
JHEP 0611 (2006) 088 [arXiv:hep-th/0608157].

\bibitem{Ahn1}
C. Ahn, 
``Meta-Stable Brane Configuration of Product Gauge Groups,''
Class.Quant.Grav.25:075001,2008 [arXiv:0704.0121].

\bibitem{Ahn3}
C. Ahn, 
``Brane Configurations for Nonsupersymmetric Meta-Stable Vacua in SQCD with Adjoint Matter,''
Class.Quant.Grav. 24 (2007) 1359-1370 [arXiv:hep-th/0608160].

\bibitem{Giveon}
A. Giveon and D. Kutasov, 
``Stable and Metastable Vacua in Brane Constructions of SQCD,'' JHEP 0802:038,2008
[arXiv:0710.1833].

\bibitem{OO}
H. Ooguri and Y. Ookouchi, 
``Landscape of Supersymmetry Breaking Vacua in Geometrically Realized Gauge Theories,''
Nucl.Phys. B755 (2006) 239-253 [arXiv:hep-th/0606061].

\bibitem{Ahn2}
C. Ahn, 
``Meta-Stable Brane Configurations by Dualizing the Two Gauge Groups,''
arXiv:0804.0051.

\bibitem{Franco}
R. Argurio, M. Bertolini, S. Franco and S. Kachru, 
``Gauge/gravity duality and meta-stable dynamical supersymmetry breaking,''
JHEP 0701 (2007) 083 [arXiv:hep-th/0610212].

\bibitem{GSW}
M. B. Green, J. H. Schwarz and E. Witten, 
``Supersting Theory,'' Cambridge University Press, 1987.

\bibitem{Barth}
W. Barth, C. Peters and A. van de Ven, 
``Compact Complex Surfaces,''
Springer-Verlag, 1984/

\bibitem{GH}
P. Griffiths and J. Harris, 
``Principles of Algebraic Geometry,''
Wiley Classics Library Edition, 1994.

\bibitem{BDO1}
E. I. Buchbinder, R. Donagi and B. A. Ovrut, 
``Vector Bundle Moduli and Small Instanton Transitions,''
JHEP 0206 (2002) 054 [arXiv:hep-th/0202084].

\bibitem{BOR}
E. I. Buchbinder, B. A. Ovrut and R. Reinbacher, 
``Instanton Moduli in String Theory,''
JHEP 0504 (2005) 008 [arXiv:hep-th/0410200].

\bibitem{Uranga}
L. E. Ibanez and A. M. Uranga, 
``Neutrino Majorana Masses from String Theory Instanton Effects,''
JHEP 0703:052,2007 [arXiv:hep-th/0609213].

\bibitem{Cvetic}
R. Blumenhagen, M. Cvetic and T. Weigand, 
``Spacetime Instanton Corrections in 4D String 
Vacua - The Seesaw Mechanism for D-Brane Models,''
Phys.Rev.D76:086002,2007 [arXiv:hep-th/0609191].

\bibitem{Shamit}
B. Florea, S. Kachru, J. McGreevy and N. Saulina, 
``Stringy Instantons and Quiver Gauge Theories,''
JHEP 0705:024,2007 [hep-th/0610003].

\bibitem{Richter}
M. Cvetic, R. Richter and T. Weigand, 
``Computation of D-brane instanton induced 
superpotential couplings - Majorana masses from string theory,''
Phys.Rev.D76:086002,2007 [arXiv:hep-th/0703028].

\bibitem{Lust}
N. Akerblom, R. Blumenhagen, D. Lust, E. Plauschinn, 
and M. Schmidt-Sommerfeld, 
``Non-perturbative SQCD Superpotentials from String Instantons,''
JHEP 0704:076,2007 [arXiv:hep-th/0612132].

\bibitem{Lerda}
R. Argurio, M. Bertolini, G. Ferretti, A. Lerda and 
C. Petersson, 
``Stringy Instantons at Orbifold Singularities,''
JHEP 06 (2007) 067 [arXiv:0704.0262].

\bibitem{Lerda1}
M. Billo, M. Frau, I. Pesando, P. Di Vecchia, A. Lerda and R. Marotta, 
``Instantons in N=2 magnetized D-brane worlds,''
JHEP 0710:091, 2007 [arXiv:0708.3806].

\bibitem{Lerda2}
M. Billo, P. Di Vecchia, M. Frau, A. Lerda, R. Marotta, I. Pesando,
``Instanton effects in N=1 brane models and the Kahler metric of twisted matter,''
JHEP 0712:051, 2007 [arXiv:0709.0245].

\bibitem{Ganor}
O. J. Ganor, 
``A Note On Zeroes Of Superpotentials In F-Theory,''
Nucl.Phys. B499 (1997) 55-66 [arXiv:hep-th/9612077].











\end{thebibliography}
\end{document}